\documentclass[journal]{IEEEtran}
\usepackage{xurl}
\usepackage{dblfloatfix}
\usepackage{tabularx}
\usepackage{wrapfig}
\usepackage{amsmath}
\usepackage{color}
\usepackage{amsmath,amsfonts,amssymb,amscd}
\usepackage{cite}
\usepackage{booktabs}
\usepackage{paralist}
\usepackage{enumitem}
\usepackage[export]{adjustbox}
\usepackage{gensymb}
\usepackage{steinmetz}
\usepackage{mathtools}
\usepackage{siunitx}
\usepackage{hyperref}
\usepackage{cuted}

\usepackage{amsmath,bm}

\usepackage[noabbrev]{cleveref}

\usepackage{mathrsfs}
\usepackage{multirow}
\usepackage{flushend}
\usepackage{graphicx}

\usepackage{algorithm2e}

\usepackage{pifont}
\newcommand{\xmark}{\ding{55}}
\newcommand{\cmark}{\ding{51}}

\begin{document}
\thispagestyle{empty}
\onecolumn
\begin{quote}
    This paper has been \textcolor{blue}{accepted} for publication in \textcolor{blue}{IEEE Transactions on Smart Grid}
\end{quote}

\vspace{1cm}
\begin{quote}
     doi: 10.1109/TSG.2024.3368277
\end{quote}
\vspace{1cm}
\begin{quote}
    Content is final as presented here, with the exception of pagination.
\end{quote}

\vspace{1cm}
\begin{quote}
    IEEE Copyright Notice: 

\vspace{0.25cm}
\noindent
\copyright 2024 IEEE. Personal use of this material is permitted.  Permission from IEEE must be obtained for all other uses, in any current or future media, including reprinting/republishing this material for advertising or promotional purposes, creating new collective works, for resale or redistribution to servers or lists, or reuse of any copyrighted component of this work in other works.
\end{quote}

\newpage
\twocolumn
\title{A Monolithic Cybersecurity Architecture for Power Electronic Systems\\
\thanks{Kirti Gupta and Bijaya Ketan Panigrahi are with the Department of Electrical Engineering, Indian Institute of Technology Delhi, New Delhi 110016, India (e-mail: \{\texttt{Kirti.Gupta, Bijaya.Ketan.Panigrahi}\}@ee.iitd.ac.in).}
\thanks{Subham Sahoo is with the Department of Energy, Aalborg University, 9220 Aalborg, Denmark (e-mail: sssa@energy.aau.dk).}
}

\author{Kirti~Gupta,
        Subham~Sahoo,~\IEEEmembership{Senior Member,~IEEE},
        and Bijaya~Ketan~Panigrahi,~\IEEEmembership{Fellow,~IEEE}
        }

\maketitle

\begin{abstract}
Power electronic systems (PES) face significant threats from various data availability and integrity attacks, significantly affecting the performance of communication networks and power system operation. As a result, several attack detection and reconstruction techniques are deployed, which makes it a costly \& complex cybersecurity operational platform with significant room for incremental extensions for mitigation against future threats. Unlike the said traditional arrangements, our paper introduces a foundational approach by establishing a monolithic cybersecurity architecture (MCA) via incorporating semantic principles into the sampling process for distributed energy resources (DERs). This unified approach concurrently compensates for the intrusion challenges posed by cyber attacks by reconstructing signals using the dynamics of the inner control layer. This reconstruction considers essential semantic attributes, like Priority, Freshness, and Relevance to ensure resilient dynamic performance. Hence, the proposed scheme promises a generalized route to concurrently tackle a global set of cyber attacks in elevating the resilience of PES. Finally, rigorous validation on a modified IEEE 69-bus distribution system and a real-world South California Edison (SCE) 47-bus network, using OPAL-RT under diverse operating conditions, underscores its robustness, model-free design capability, scalability, and adaptability to dynamic cyber graphs and system reconfiguration. 
\end{abstract}

\begin{IEEEkeywords}
Data dropout, distributed control, false-data injection attacks (FDIAs), inner control loop dynamics, latency attack, power electronic systems (PES), semantic sampling, time synchronization attack (TSA).
\end{IEEEkeywords}

\IEEEpeerreviewmaketitle

\section{Introduction}
\IEEEPARstart{I}{n} the realm of power electronic systems (PES), the integration of control and communication is pivotal for achieving efficiency, sustainability, and flexibility. However, conventional centralized secondary controllers (SCs) within PES exhibit inherent limitations, including stringent communication bandwidth requirements, susceptibility to single-point failures, and intricate computational demands. Addressing these constraints, the adoption of a distributed secondary control (DSC) architecture has emerged as a compelling alternative, harnessing insights from neighboring agents \cite{ref1}.  Ensuring synchronization among network agents entails the utilization of global navigation satellite signals (GNSSs), acclaimed for their global reach and precision \cite{ref2}. GNSS facilitates the transfer of time information among intelligent electronic devices (IEDs) and merging units through mechanisms like precision time protocol (PTP), inter-range instrumentation group time code B (IRIG-B), or one pulse per second (1PPS) over coaxial cable \cite{ref3}.

However, the integration of communication networks into PES introduces several bottlenecks including delays, data losses, and uncertain connections \cite{ref4}. These factors can potentially result in delayed transmission of control signals among distributed energy resources (DERs), consequently impairing the overall system performance. Furthermore, these integrated cyber-physical systems are also prone to cyber attacks \cite{ref5,ref6}. Given PES's characteristics of low system inertia and swift responsiveness, such attacks can have more profound implications compared to bulk power systems. This research is dedicated to mitigating data availability attacks \cite{ref7} and data integrity attacks \cite{ref8}, which can be deliberately introduced by malicious entities. Data availability attacks encompass latency attacks, data dropouts, and time-synchronization attacks (TSAs), all of which compromise the timeliness and reliability of signal exchanges. Data integrity attacks, often referred to as false data injection attacks (FDIAs), manipulate transmitted data, significantly disrupting system control.

\begin{figure*}[t!]
        \centering
	    \includegraphics[clip, trim=0.5cm 2.5cm 5cm 0.7cm, width=0.7\linewidth]{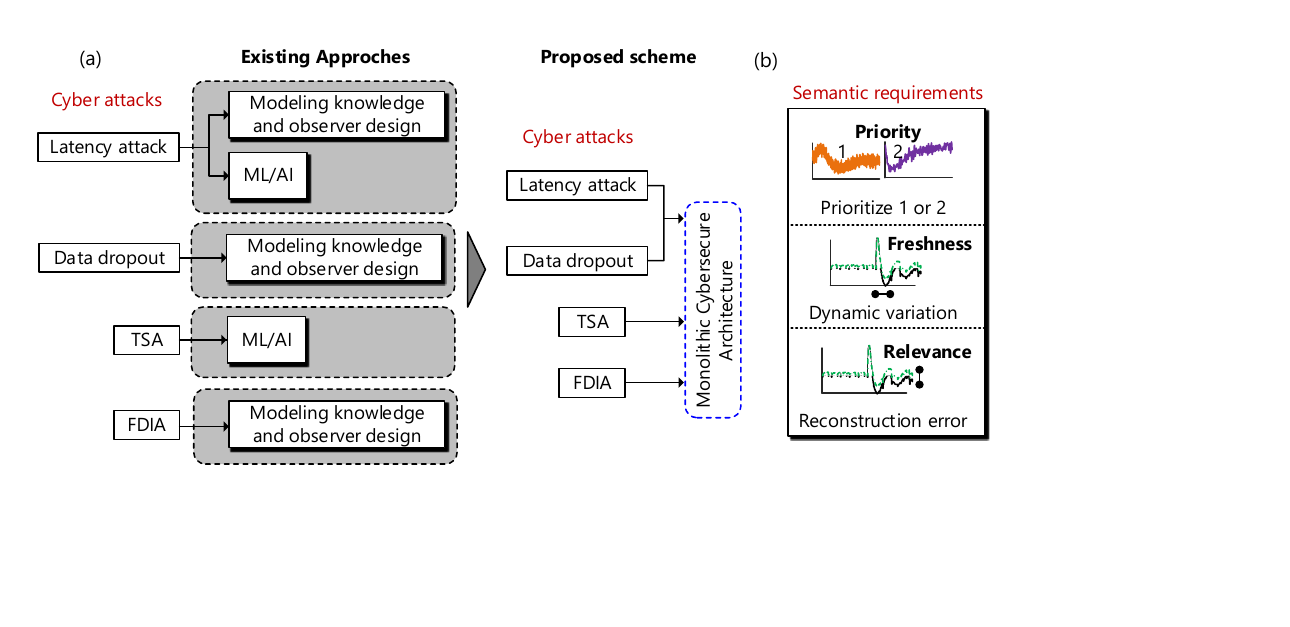}
	    \caption{\textbf{(a)} Paradigm shift from a complex architecture of cyber-physical resiliency methods to a monolithic cybersecurity architecture (MCA) -- the proposed scheme with semantic sampling offers unified resiliency against data availability and data integrity attacks, \textbf{(b)} Semantic requirements governing the proposed MCA scheme.}
	    \label{fig:Semantics}
\end{figure*}
The development of robust controllers, capable of functioning within real-time constraints to withstand adversarial attacks, is of utmost significance. Diverse methodologies have been presented, each carrying inherent limitations. To make the system resilient against data availability attacks, the strategy outlined in \cite{ref9} concentrates on enhancing microgrid communication networks, while \cite{ref10} integrates a controller with a supply-demand optimization algorithm. Despite their advantages, these approaches may impose substantial computational burdens in larger networks. Their susceptibility to initial conditions and hyperparameters may result in sub-optimal outcomes. Cooperative control, leveraging predictive control theory as proposed in \cite{ref11}, presents potential challenges in handling unknown dynamics. This may possibly lead to performance degradation or instability during data availability attacks. TSA detection scheme \cite{ref12} and FDIA detection \cite{ref12a}, focus solely on attack detection and necessitate a training phase to comprehend the system's normal behavior. Relying heavily on extensive training data introduces complexities such as increased memory usage, substantial data prerequisites, and privacy concerns. These constraints, coupled with the inability to explain abnormal data variations, diminish the suitability of data-driven approaches for power electronic converter applications. The existing approaches developed against data availability attacks \cite{ref13a} and data integrity attacks \cite{ref13b} mandate multiple observers/estimators, resulting in increased computational demands and complexity. Earlier studies have separately investigated data availability attacks and data integrity attacks, yielding various detection and control methodologies thereby increasing the computational complexity. 
    
    To fill this gap, this paper introduces a foundational approach that leverages semantic principles to establish a monolithic cybersecurity architecture (MCA), specifically designed for PES. Although the infusion of semantic principles into PES may seem like a recent initiative, it is noteworthy that the concept of \textit{semantics} has already found significant applications across various domains. In the broader landscape of communication systems, semantic approaches have proven effective \cite{B1}. Likewise, in the domain of networked intelligent systems, semantic techniques have been employed to augment system intelligence and facilitate meaningful interaction \cite{B2}. A notable example is in the field of speech recognition, where semantics play a crucial role in transforming spoken language into written text, enabling more precise and efficient speech-to-text transmission \cite{B3}. Additionally, as the world transitions into the era of post-5G wireless connectivity, semantic considerations are instrumental in shaping the next generation of wireless networks \cite{B4}. For a comprehensive exploration of semantic communication and its applications, readers are encouraged to delve into the detailed insights provided in \cite{B5}.
    
    The presented methodology employs judicious risk-aware sampling to systematically reconstruct and disseminate critical information within PES, ensuring its precise utilization. This study introduces the innovative concept of semantic communication (SemCom) and employs advanced sampling techniques, prioritizing the filtration of the most significant events through signal processing methods. Upholding privacy as a fundamental aspect of ethical research, our proposed MCA is strategically integrated into each DER. This distributed approach not only empowers individual DERs with localized intelligence regarding cyber attacks, but also adeptly navigates the privacy challenges associated with handling sensitive data. The principal contributions of this work are summarized as:

\begin{itemize}
    \item \textbf{Innovative Architecture:} The proposed MCA leverages vital information derived from the inner control loop dynamics. This enables the generation of reconstructed signals for local SC, addressing both data availability and data integrity attacks. This strategic approach significantly reduces redundant data transmissions. 
    \item \textbf{Robustness Against Cyber Attacks:} The proposed scheme exhibits resilience against various cyber threats, including latency attacks, data dropouts, TSAs, and FDIAs. It also ensures that SC objectives are consistently met even in the presence of such adversarial scenarios. 
    \item \textbf{Distributed Approach:} The proposed MCA adopts a distributed approach, offering a streamlined alternative to intricate centralized methods. This approach empowers individual DERs to autonomously manage local cyber attack compensation, enhancing operational efficiency and manageability.
    \item \textbf{Model-Agnostic Design:} Unlike conventional methods relying on device-specific models, our proposed scheme is model-agnostic and can be generalized for the most prominent cyber attacks. This simplifies implementation and saves on computational resources, eliminating the need for a multitude of device-specific models and promoting broader applicability.
    \item \textbf{Training-Free Operation:} In contrast to resource-intensive training-based approaches demanding extensive computational resources, large datasets, and meticulous hyperparameter tuning, our approach operates without the need for training. Additionally, it imposes no extra hardware requirements, offering a more practical and resource-efficient solution.
\end{itemize}

The structure of this paper is as follows: Section II delves into the principle of semantics and its impact on cybersecurity. Section III provides an overview of the modeling aspects pertaining to cyber-physical PES. The challenges related to data availability and data integrity attacks are detailed in Section IV. The novel MCA framework is expounded in Section V. Section VI outlines the real-time simulation testbed setup and rigorously evaluates the performance of the proposed scheme. Finally, Section VII concludes the paper by summarizing the findings and outlining potential future directions.

\section{Semantics -- Principles \& Impact on Cybersecurity}
As highlighted in the existing literature, prevailing methodologies have primarily directed their efforts towards addressing data availability attacks and data integrity attacks as distinct challenges, as illustrated in Fig. \ref{fig:Semantics}(a). However, these approaches exhibit inherent constraints, notably necessitating multiple detection metrics (e.g., $\mathrm{D}_1$ and $\mathrm{D}_2$ in Fig. \ref{fig:Semantics}) and observers/estimators for each attack scenario, thereby escalating the computational resources. Consequently, this study introduces a holistic strategy confronting both types of attacks while concurrently mitigating communication delays. 

Drawing inspiration from the field of linguistics, there has been a notable surge of interest within the communications theoretical community towards the concept of SemCom \cite{ref14}. SemCom transcends Shannon's conventional viewpoint of reliable transmission, which treats all bits uniformly. Instead, SemCom focuses on the rationale behind data transmission, emphasizing the interpretation of content, context, and timing aspects at the receiver's end \cite{ref15}. This perspective provides methodologies that aim to reduce unnecessary transmissions and prioritize data, thereby enhancing overall performance. 

\begin{figure}[h!]
        \centering
	    \includegraphics[clip, trim=0.5cm 3.9cm 3.9cm 0.7cm, width=1\linewidth]{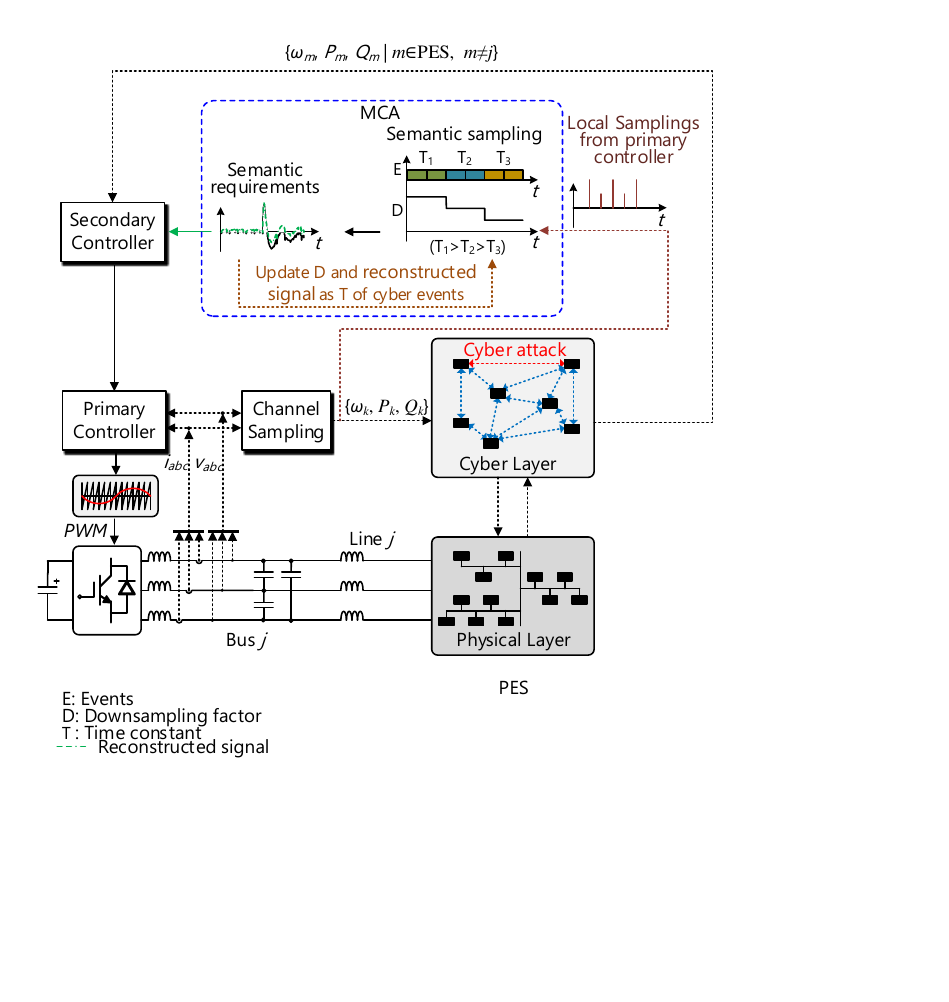}
	    \caption{Schematic diagram of the deployment of the proposed scheme (enclosed in dotted blue box) in the existing PES. The semantic information exchange from local error measurements drives the estimation and reconstruction process (to SC) during cyber attacks.}
	    \label{fig:GSCP4}
\end{figure}

In the context of semantic sampling, the evaluation of information criticality or significance revolves around the following three key attributes (in Fig. \ref{fig:Semantics}(b)):
\begin{enumerate}
\item \textbf{Value} of information (VoI), establishes the priority of information. This ensures that the significant data reaches the right destination for efficient processing and decision-making. 

\item \textbf{Freshness} is the concept of delivering new updates at the right time. It is quantified as $\mathrm{F(t)}=\mathrm{t}-\mathrm{S(t)}$, denoting the time discrepancy between the present time and the arrival of the most recent packet. This quantification determines the age of information (AoI).

\item \textbf{Relevance} quantifies the magnitude of change in a process since the last recorded sample, to capture relevant updates and meaningful insights. Furthermore, the requisite alteration is precisely conveyed to the right place.
\end{enumerate}

Leveraging the aforementioned semantic requirements, the error signals from the inner voltage control (VC) loop are downsampled (D is the downsampling factor) and the reconstructed signals are sent to each SC for attack compensation to each DER within PES, as depicted in Fig. \ref{fig:GSCP4}. The signals continuously undergo downsampling according to event time constants (T), which determine events' significance. This minimizes redundant bulk data transmissions, thus improving overall efficiency of the system. Therefore, we only utilize the most significant data at the right time at the right place. This approach prioritizes the preservation of estimation accuracy under attack scenarios, such as latency attacks, data dropouts, TSAs and FDIAs. Furthermore, the semantic attributes, namely priority, freshness, and relevance, are determined by factors such as prioritization of the most significant local signal used for estimation, dynamic variation, and the reconstruction error, respectively. Consequently, the integration of semantic models establishes a standardized mechanism for interpreting relevant data collected from diverse devices and sensors across the network, fostering improved decision-making processes and enhancing overall system performance. 

\section{Modeling Preliminaries}
\subsection{Physical Framework}
\begin{figure}[h!]
        \centering
	     \includegraphics[clip, trim=0.5cm 2.8cm 4.2cm 0.5cm, width=1\linewidth]{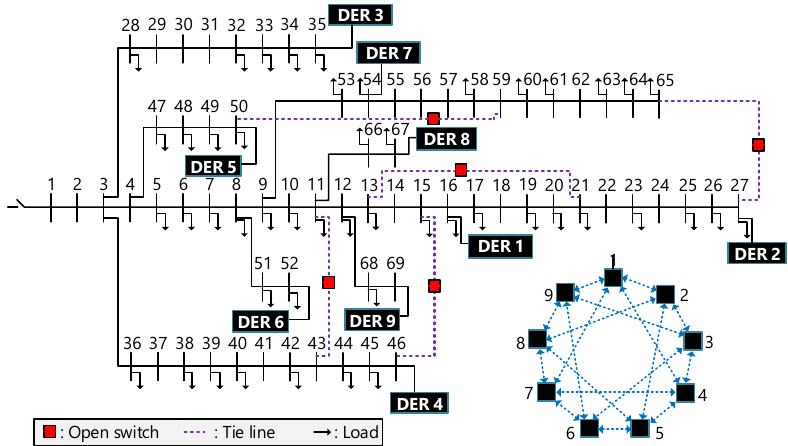}
	    \caption{Modified 69-bus distribution system with nine DERs in partially-connected topology \cite{ref16}.}
	    \label{fig:69bus}
\end{figure}

\begin{figure}[h!]
        \centering
	     \includegraphics[clip, trim=0.5cm 2.5cm 3.5cm 0.5cm, width=0.9\linewidth]{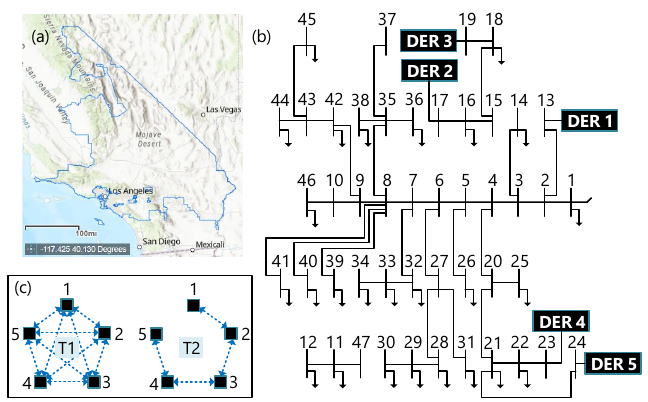}
	    \caption{\textbf{(a)} South California Edison's (SCE's) Distributed Energy Resource Interconnection Map (DERiM) \cite{ref18}; \textbf{(b)} real-world SCE 47-bus network with five DERs \cite{ref17} and; \textbf{(c)} cyber topologies T1 and T2.}
	    \label{fig:Map}
\end{figure}

This study examines two test PES: a modified IEEE 69-bus system with nine DERs \cite{ref16} in Fig. \ref{fig:69bus} and a real-world South California Edison (SCE) 47-bus network with five DERs \cite{ref17} in Fig. \ref{fig:Map}. To depict the modeling and control structure of a PES, each DER comprises of a DC source (indicating an energy storage system), a DC/AC converter (inverter), LC filter, and RL output impedance, as shown in Fig. \ref{fig:GSCP4}. The $dq$ axis control framework encompasses internal voltage control (VC) and current control (CC) loops, coupled with the primary droop control (DC) loop, inside the primary controller. Synchronized measurements, enabled by global positioning system (GPS), are transmitted to these controllers to enable their operation.

The frequency and voltage droop schemes for PES are:

\vspace{-0.1cm}
\begin{equation}
    \label{EQN1}
    {{\omega }_{j}^*}(t)=\omega_{\mathrm{nom}}-\mathrm{m}^{\mathrm{p}}_{j}{\mathrm{P}_{j}}\mathrm{(t)}
\end{equation}
\begin{equation}
    \label{EQN2}
    \mathrm{V}^{\mathrm{d*}}_{j}\mathrm{(t)}=\mathrm{V_{nom}}-\mathrm{n}^{\mathrm{q}}_{j}{\mathrm{Q}_{j}}\mathrm{(t)}\hspace{0.2cm},\hspace{0.2cm} \mathrm{V}^{\mathrm{q*}}_{j}\mathrm{(t)}=0
\end{equation}
Here, the subscript '$j$' indicates parameters related to the $j^{th}$ DER. The terms $\omega_{\mathrm{nom}}$ and $\mathrm{V}_{\mathrm{nom}}$ represent the nominal frequency and voltage of the AC system, respectively. The local reference frequency and voltage of a DER are designated as ${{\omega }_{j}^*}$ and $\mathrm{V}^{\mathrm{dq*}}_{j}$, where $\mathrm{V}^{\mathrm{dq*}}_{j}\mathrm{(t)}=[\mathrm{V}^{\mathrm{d*}}_{j}\mathrm{(t)} \hspace{0.2cm} \mathrm{V}^{\mathrm{q*}}_{j}\mathrm{(t)}]^{\mathrm{T}}$. Moreover, the active power droop coefficient is represented by $\mathrm{m}^{\mathrm{p}}$, and the reactive power droop coefficient is denoted as $\mathrm{n}^{\mathrm{q}}$. For a comprehensive understanding of the control layer modeling, refer to \cite{ref19}. As the primary control inherently leads to non-zero steady-state error, the integration of the DSC scheme, depicted in Fig. \ref{fig:GSCP4}, is introduced and discussed in forthcoming subsection. Such framework aids to counteract these errors, thereby enhancing overall system performance.

\subsection{Cyber Framework}
Consider DERs in a sparsely-connected DSC architecture denoted as a PES with `$M$' agents (or nodes in the cyber layer), represented as $\textbf{x} = \{\mathrm{x}_1, \mathrm{x}_2, … , \mathrm{x}_M\}$. These agents are interconnected by edges $\mathbf{E}$ through an adjacency matrix, ${\mathbf A_\text{G}} = [\mathrm{e}_{jm}]\in{\textbf{R}^{N\times{N}}}$. The neighbors of the $j^{th}$ agent are denoted as, $\mathrm{N}_{j} = \{ {m} \ | \ (\mathrm{x}_m, \mathrm{x}_j) \in \mathbf{E} \} $. The communication weight $\mathrm{e}_{jm}$ (from agent $m$ to agent $j$) is modeled as: $\mathrm{e}_{jm} >$ 0, if ($\mathrm{x}_j$, $\mathrm{x}_m$) $\in$ $\mathbf{E}$. If there is no cyber link between $\mathrm{x}_j$ and $\mathrm{x}_m$, then $\mathrm{e}_{jm}$ = 0. Each agent exchanges information with neighboring agent(s), $\mathrm{\Psi}_m=[\mathrm{\omega}_m \hspace{0.2cm} \mathrm{m}_m^{\mathrm{p}}\mathrm{P} \hspace{0.2cm} \mathrm{n}_m^{\mathrm{q}}\mathrm{Q}]^\mathrm{T}$. The incoming information matrix is, $\mathbf{D}_\text{in} = \texttt{diag}\{\mathrm{d}_j^{\mathrm{in}}\}$, where $\mathrm{d}_j^{\mathrm{in}}=\sum_{m\in \mathrm{N}_{j}}\mathrm{e}_{jm}$. The Laplacian matrix $\mathbf{L}$ = [$\mathrm{l}_{jm}$], where $\mathrm{l}_{jm}$ are its elements defined such that, $\mathbf{L}$ = ${\mathbf{D}_\text{in}} – {\mathbf{A}_\text{G}}$. Referring to \cite{ref19}, the local reference frequency and voltage of a DER, initially expressed in \eqref{EQN1} and \eqref{EQN2}, are re-defined as:
\vspace{-0.1cm}
\begin{equation}
    \label{EQN3}
    {{\omega }_{j}^*}\mathrm{(t)}=\omega_{\mathrm{nom}}-\mathrm{m}^{\mathrm{p}}_{j}{\mathrm{P}_{j}}\mathrm{(t)}+\delta\omega_{j}\mathrm{(t)}
\end{equation}
\begin{equation}
    \label{EQN4}   \mathrm{V}_{j}^{\mathrm{d*}}\mathrm{(t)}=\mathrm{V}_{\mathrm{nom}}-\mathrm{n}^{\mathrm{q}}_{j}{\mathrm{Q}_{j}}\mathrm{(t)}+\delta \mathrm{V}_{j}\mathrm{(t)}\hspace{0.2cm}
\end{equation}
Here, $\delta {\mathrm{\omega}}$ and $\delta {\mathrm{V}}$ are the frequency and voltage correction terms from the SC, that can be given by: 
\vspace{-0.1cm}
\begin{equation}
\label{EQN5}
\begin{aligned}
    \delta \mathrm{\omega}_{j}\mathrm{(t)}=&-\mathrm{H}_1(s)[\mathrm{\omega_{nom}}-\omega_j\mathrm{(t)}+\\ &\mathrm{c}_j\sum\limits_{m\in {\mathrm{N}_{j}}}{{\mathrm{e}_{jm}}\left( \mathrm{\omega}_m\mathrm{(t)}-\mathrm{\omega}_j\mathrm{(t)} \right)}+\\ &\mathrm{c}_j\sum\limits_{m\in {\mathrm{N}_{j}}}{{\mathrm{e}_{jm}}\left(\mathrm{m}_m^\mathrm{p}\mathrm{P}_m\mathrm{(t)}-\mathrm{m}_j^\mathrm{p}\mathrm{P}_j\mathrm{(t)} \right)}]
\end{aligned}
\end{equation}
\vspace{-0.1cm}
Similarly,
\begin{equation}
\label{EQN6}
    \delta \mathrm{V}_{j}\mathrm{(t)}=-\mathrm{H_2(s)}[\mathrm{c}_j\sum\limits_{m\in {\mathrm{N}_{j}}}{{\mathrm{e}_{jm}}\left(\mathrm{n}_m^\mathrm{q}\mathrm{Q}_m\mathrm{(t)}-\mathrm{n}_j^\mathrm{q}\mathrm{Q}_j\mathrm{(t)} \right)}]
\end{equation}
where, $\mathrm{H_1(s)}$ and $\mathrm{H_2(s)}$ denote the PI controllers for frequency restoration and proportional active power sharing, as well as proportional reactive power sharing, respectively. The local control input of the SC can be defined as:
\vspace{-0.1cm}
\begin{equation}
\label{EQN7}
    \bm{\zeta}_j\mathrm{(t)}=\mathrm{c}_j\sum\limits_{m\in {\mathrm{N}_{j}}}\underbrace{{{\mathrm{e}_{jm}}\left( \boldsymbol{\Psi}_m\mathrm{(t)}-\boldsymbol{\Psi}_j\mathrm{(t)} \right)}}_{\bm{\varrho}_{jm}\mathrm{(t)}}
\end{equation}
where, $\bm{\zeta}_j=[\mathrm{\zeta}_j^\mathrm{p} \hspace{0.2cm} \mathrm{\zeta}_j^\mathrm{q}]$, $\bm{\varrho}_{jm}=[\mathrm{\varrho}_{jm}^\mathrm{p} \hspace{0.2cm} \mathrm{\varrho}_{jm}^\mathrm{q}]$, depending on the elements in $\boldsymbol{\Psi}$; and $\mathrm{c}_j$ is the convergence parameter. 

These information exchanges may encounter limitations caused by data availability and data integrity cyber attacks. These challenges can disrupt the system's monitoring and controllability, leading to the absence and corruption of critical information, as elaborated in the following section.

\section{Modeling of Cyber Attacks}

\subsection{Latency Attacks and Data Dropouts}
The effectiveness and reliability of the DSC architecture heavily rely on real-time periodic communication. Nonetheless, the occurrence of congestion and delays in data packets can usher in unanticipated disruptions. These delays, influenced by factors like cyber sampling rate and data volume, can span from milliseconds to seconds. If these delays surpass the operational time limits of the SC, missed updates can incite oscillatory instability and instigate system failure \cite{ref20}. 

Moreover, malicious cyber attackers can exploit these vulnerabilities by intentionally introducing time delays to critical messages, known as latency attacks. Such attacks can severely compromise system functionality, particularly when time-sensitive information exchange occurs between SCs. Additionally, network congestion can cause frequent data dropouts, further undermining communication and dynamic performance.

\textbf{Attack model:}
When confronted with latency attacks, the control laws that govern the behavior of cyber-physical PES can undergo substantial modifications. These alterations in the SC can be mathematically represented as follows:
\vspace{-0.1cm}
\begin{equation}
\label{EQN8}
    \bm{\zeta}_j(t)=\mathrm{c}_j\sum\limits_{m\in {\mathrm{N}_{j}}}{{\mathrm{e}_{jm}}\left( \boldsymbol{\Psi}_m(\mathrm{t}-\mathrm{\tau}_m)-\boldsymbol{\Psi}_j(\mathrm{t}-\mathrm{\tau}_j) \right)}
\end{equation}
where, $\mathrm{\tau}_j$ represents the delay from the local unit, while $\mathrm{\tau}_m$ represents the delay from neighboring units. By utilizing the Leibnitz formula, the delayed parameter can be expressed as $\mathrm{\Psi(t-\tau)=\Psi(t)-\int_{t-\tau}^{t}\dot{\Psi}(s)ds}$. Substituting this expression into equation \eqref{EQN8} for a delay of $\mathrm{\tau}_m$, we obtain $\mathrm{\dot{\Psi}(t)=-\textbf{L}\Psi(t)-\textbf{A}\int_{t-\tau_m}^{t}\dot{\Psi}(s)ds}$. A similar expression can be derived for the local delay. For a fixed, undirected, and connected cyber graph, equilibrium is achieved if and only if $0<\mathrm{\tau}<\mathrm{\frac{\pi}{2\lambda_{max}\mathbf{L}}}$, where $\lambda_{max}$ represents the largest eigenvalue of the matrix $\mathbf{L}$. Thus, the communication delay ($\mathrm{\tau}$) must be constrained within these limits in order to obtain $\mathrm{\dot{\Psi}(t)=0}$.

\begin{figure}[h!]
        \centering
	    \includegraphics[clip, trim=0.5cm 2cm 4cm 0.7cm,width=1\linewidth]{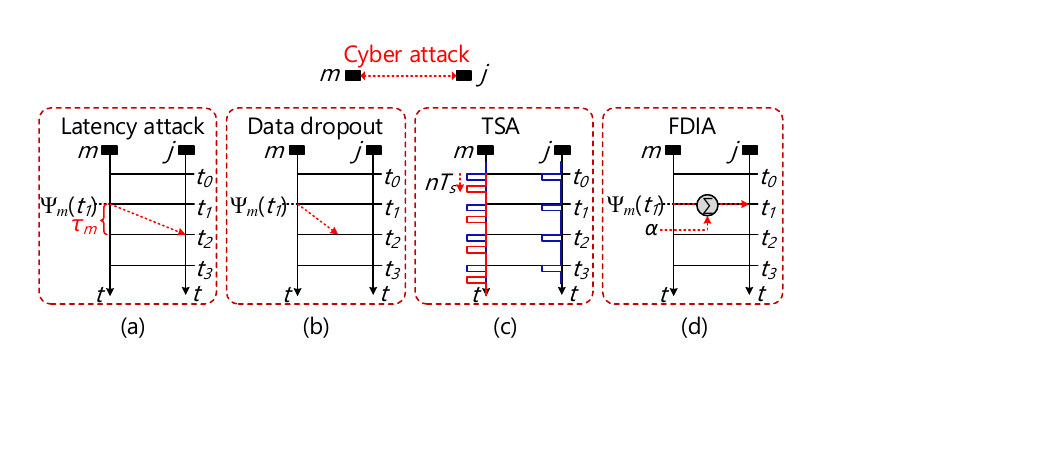}
	    \caption{The depicted models illustrate cyber attacks on two communicating SCs in terms of \textbf{(a)} latency attack; \textbf{(b)} data dropout; \textbf{(c)} TSA; and \textbf{(d)} FDIA.}
     \label{fig:AttackN}
\end{figure}

\subsection{Time Synchronization Attacks (TSAs)}
Time-synchronized real-time measurements in cyber-physical networks hold paramount importance, rendering them susceptible to malicious attacks. In recent times, there has been a notable upswing in TSAs, which has raised significant concerns across diverse sectors. This issue becomes particularly critical due to potential vulnerabilities in the integrity of GPS signals. Unintentional sources like radio frequency (RF) interference and space weather events, such as solar flares,  hold the potential to introduce timing inaccuracies or even outright signal loss. These discrepancies bear substantial risks for applications hinging on precise temporal synchronization.

Furthermore, GPS receivers in substation clocks or merging units are vulnerable to deliberate attacks orchestrated by malicious entities. A notable example is the receiver-spoofer mechanism \cite{ref21}, wherein the spoofer masquerades as a GPS receiver, manipulates authentic GPS signals, and transmits spoofed signals to the target. Both space-based time synchronization (SBTS) and network-based time synchronization (NBTS) mechanisms \cite{ref22} lack integrated security controls, rendering them highly susceptible to TSAs.

\textbf{Attack model:} In the case of a TSA, where the time-stamped information of $\boldsymbol{\Psi}_m\mathrm{(t)}$ is falsified by $\mathrm{nT_s}$ samples, the resulting information can be expressed as:
\vspace{-0.1cm}
\begin{equation}
\label{EQN9}
    \Psi_m^{\mathrm{T}}\mathrm{(t)}=\Psi_m\mathrm{(t \pm nT_s)}
\end{equation}

\subsection{False Data Injection Attacks (FDIAs)}
Adversaries possess the capability to manipulate the data interchanged between agents, strategically aiming to deceive the control system into pursuing actions that serve the attacker's agenda. Such actions may include destabilizing the PES, inducing power imbalances, or causing harm to the system's components. It can be categorized as balanced attack and unbalanced attack \cite{ref23}, each having distinct modeling implications. Balanced attacks ensure a stable and feasible solution, aligning with the SC objectives. In contrast, unbalanced attacks deviate from SC objectives.

\textbf{Attack model:} In the event of a FDIA, where the exchanged information between neighbors  ($\Psi_m\mathrm{(t)}$) is fabricated with $\mathrm{\alpha}=[\mathrm{\alpha^\omega} \hspace{0.2cm} \mathrm{\alpha}^\mathrm{P} \hspace{0.2cm} \mathrm{\alpha}^\mathrm{Q}]^\mathrm{T}$, the resultant attacked information can be mathematically represented as:

\vspace{-0.1cm}
\begin{equation}
\label{EQN10}
\Psi_m^{\mathrm{F}}\mathrm{(t)}=\Psi_m\mathrm{(t)}+ \Lambda. \alpha
\end{equation}
where, $\Lambda$ is a binary number signifying `$\mathrm{1}$' for presence of attack and `$\mathrm{0}$' for no attack. Furthermore, these attacks can also be executed in a coordinated manner to deceive the system operator as follows:

\vspace{-0.3cm}
\begin{equation}
\label{EQN11}
\dot{\Psi}\mathrm{(t)}=-(\mathbf{L}\Psi\mathrm{(t)}+\alpha)
\end{equation}
where, $\Psi$ and $\alpha$ denote column matrices of the measured information and the attack signal respectively. Therefore, \eqref{EQN11} can be used to model the balanced and unbalanced attacks as:

\begin{equation}
   \dot{\Psi}\mathrm{(t)} \left\{\begin{aligned}
& =
  0 \hspace{0.2cm}; \text{balanced attack}\\
& \begin{aligned}
  \neq
  0 \hspace{0.2cm}; \text{unbalanced attack} \\
  \end{aligned}
\end{aligned}\right.
\end{equation}
 
Further details on FDIA modeling can be found in \cite{ref23}. In the domain of PES, the aforementioned cyber attacks carry significant implications for the system's optimal functioning. These attacks, illustrated in Fig. \ref{fig:AttackN}, can trigger a spectrum of issues, ranging from sub-optimal operational conditions to system instability. Such scenarios might lead to unintended disconnections of sources and/or loads, ultimately resulting in partial or complete shutdown of the system, thereby undermining the reliability of power supply. To effectively tackle these challenges, the establishment of a one-stop simple monolithic solution becomes imperative. This architecture must be capable of addressing unforeseen delays, data dropouts, and cyber attacks. With this objective in mind, this paper embarks on the development of a novel MCA framework, which is comprehensively explained in the next section.

\section{Proposed MCA Framework}
In the pursuit of comprehending the role of information in decision-making during instances of data availability (and/or) data integrity attacks in networked PES, this study introduces the concept of information semantics. Through this integration, it aims to provide a more nuanced and comprehensive insight into the functioning of information under such circumstances.

In the context of PES, semantic contextuality entails capturing the attributes inherent to signals within the inner control loops of each DER, such as their timeliness and intrinsic value. This enables the reconstruction of vital information that remains critical for compensation strategies in scenarios involving random delay attacks and FDIAs, as visually depicted in Fig. \ref{fig:GSCP4}. Particularly, concerning the distributed control of AC distribution systems, prompt consensus negotiations among agents stand as a crucial factor for achieving global frequency regulation and proportional active/reactive power sharing. Consequently, a semantic-aware transmission approach that respects the time-varying value of signals becomes indispensable for realizing the objectives of the SCs. The architecture entails a series of distinct steps, presented as:

\renewcommand{\theenumi}{\roman{enumi}}%
\begin{enumerate}
  \item \textit{Risk-aware semantics:} To grasp the intricacies of the proposed methodology, it becomes imperative to apply the PI consensusability law \cite{ref24} to anticipate the inherent semantics at the physical layer. This is accomplished by investigating the response of each control loop when subjected to disturbances. It is of significance to acknowledge that the introduced schemes at the local level target individual SC units. Implementing the proposed framework entails a systematic approach, commencing with the extraction of crucial information from the error signal associated with the VC loop. This particular error signal is denoted as $\bm{\varrho}_j^{\textbf{dqVC}}\mathrm{(t)}$, where $\bm{\varrho}_j^{\textbf{dqVC}}\mathrm{(t)}=[\mathrm{\varrho}_j^{\mathrm{dVC}}\mathrm{(t)} \hspace{0.2cm} \mathrm{\varrho}_j^{\mathrm{qVC}}\mathrm{(t)}]^{\mathrm{T}}$.\\
  
  \item \textit{Semantic sampling:} Next, the signal $\bm{\varrho}_j^{\textbf{dqVC}}\mathrm{(t)}$ undergoes downsampling, resulting in a reduced rate of data acquisition. This downsampling process is implemented as:
  \vspace{-0.1cm}
\begin{equation}
\label{EQN12}
    \mathrm{\varrho}_j^\mathrm{dD}=\sum_{w=0}^{\mathrm{W-1}}\mathrm{\varrho}_j^{\mathrm{dVC}}[\mathrm{nD}-w].\delta[w]
\end{equation}
\vspace{-0.1cm}
\begin{equation}
\label{EQN13}
    \mathrm{\varrho}_j^\mathrm{qD}=\sum_{w=0}^{\mathrm{W-1}}\mathrm{\varrho}_j^{\mathrm{qVC}}[\mathrm{nD}-w].\delta[w]
\end{equation}
In this context, the downsampling procedure entails the utilization of an impulse response, indicated as $\delta[w]$, characterized by a window length of W and a downsampling factor of D. Downsampling, conventionally used as a resampling mechanism to mitigate memory consumption and curtail the resolution of the incoming signal, is repurposed in this study. Here, the downsampling operation aims to align the dynamic behavior of the error signals supplied to the VC loop with the error being fed to the SC, as shown in the Fig. \ref{fig:GSCP4}. This alignment is crucial for achieving synchronization of the  multi-time scale error signals, which is achieved by sampling based on the time constant of events. Moreover, it facilitates efficient utilization of bandwidth resources, contributing to overall system optimization. \\
\RestyleAlgo{ruled}
\SetKwComment{Comment}{/* }{ */}

\begin{algorithm}[h!]
\caption{Proposed MCA at $j^{th}$ DER}\label{alg:two}
\textbf{Inputs:}  Error signals to inner VC loop ($\bm{\varrho}_j^{\textbf{dqVC}}\mathrm{(t)}$), window length (W), downsampling factor (D), local control inputs to SC ($\bm{\zeta}_j^{\textbf{pq}}\mathrm{(t)}$), tunable parameter ($\beta$), controller time constant of $\mathrm{H_1(s)}$ and $\mathrm{H_2(s)}$ PI control loops ($\mathrm{T_c=K_p/K_i}$), triggering moment ($\mathrm{t}_a$), tunable gains ($\mathrm{g_1}$ and $\mathrm{g_2}$) \\

\textbf{Signals:} impulse response ($\delta[w]$), downsampled signal ($\bm{\varrho}_j^{\textbf{dqD}}\mathrm{(t)}$), error fed to semantic prediction policy ($\bm{\varrho}_j(\mathrm{t}_a)$), reconstructed signals ($\bm{\varrho}_j^{\textbf{R}}(\mathrm{t}_a)$), final predictive inputs to SC ($\bm{\zeta}_j^{\textbf{pqf}}\mathrm{(t)}$), freshness ($\mathrm{F(t)}$), relevance ($\mathrm{R(t)}$), $\mathrm{S(t)}$ is the timestamp of the latest packet received at destination by time $\mathrm{t}$. \\

\textbf{Note:} $\bm{\varrho}_j^{\textbf{dqVC}}\mathrm{(t)}=[\mathrm{\varrho}_j^{\mathrm{dVC}}\mathrm{(t)} \hspace{0.2cm} \mathrm{\varrho}_j^{\mathrm{qVC}}\mathrm{(t)}]^{\mathrm{T}}$, $\bm{\varrho}_j^{\textbf{dqD}}\mathrm{(t)}=[\mathrm{\varrho}_j^\mathrm{dD}(t) \hspace{0.2cm} \mathrm{\varrho}_j^\mathrm{qD}(t)]^{\mathrm{T}}$,
$\bm{\zeta}_j\mathrm{(t)}=[\mathrm{\zeta}_j^\mathrm{p}\mathrm{(t)} \hspace{0.2cm} \mathrm{\zeta}_j^\mathrm{q}\mathrm{(t)}]^{\mathrm{T}}$,
$\bm{\varrho}_j(\mathrm{t}_a)=[{\mathrm{\varrho}_j^\mathrm{p}(\mathrm{t}_a) \hspace{0.2cm} \mathrm{\varrho}_j^\mathrm{q}(\mathrm{t}_a)}]^{\mathrm{T}}$,
$\bm{\varrho}_j^{\textbf{R}}(\mathrm{t}_a)=[\mathrm{\varrho}_j^{\mathrm{Rp}}(\mathrm{t}_a) \hspace{0.2cm} \mathrm{\varrho}_j^{\mathrm{Rq}}(\mathrm{t}_a)]^{\mathrm{T}}$,
$\bm{\zeta}_j^{\textbf{pqf}}\mathrm{(t)}=[\mathrm{\zeta}_j^{\mathrm{pf}}\mathrm{(t)} \hspace{0.2cm} \mathrm{\zeta}_j^{\mathrm{qf}}\mathrm{(t)}]^{\mathrm{T}}$ \\
\vspace{0.5cm}
\KwData{$k \geq 0$}
$K \gets k$\;
// Initialize: $\mathrm{F(t)} = 0$, $\mathrm{R(t)} \neq 0$ \\ 
\While{$K \neq 0$}{
\vspace{0.1cm}
\eIf{($\mathrm{F(t)} = 0 \hspace{0.2cm} \&\& \hspace{0.2cm} \mathrm{R(t)} \neq 0$)}{
    \vspace{0.1cm}
    // Compute freshness \eqref{EQ14}: $\mathrm{F(t)}=\mathrm{t}-\mathrm{S(t)}$ \\
    \vspace{0.1cm}
    // Semantic sampling: \eqref{EQN12} and \eqref{EQN13} \\
    // Triggers generation with the semantic prediction policy condition \\
  \eIf{(\eqref{EQN15} holds)}{
    \vspace{0.1cm}
    // Reconstruction of signals \\
    $\bm{\varrho}_j^{\textbf{R}}(r\mathrm{t}_a+\Gamma)$=$\bm{\varrho}_j(r\mathrm{t}_a)$; 0$\leq \Gamma < \mathrm{t}_a$ and r=0,1,2,...\\
    \vspace{0.1cm}
    // Reconstructed signals fed back to SC with tunable gains by \eqref{EQN16} \\
    
    \vspace{0.1cm}
    // Final predictive inputs fed to the SC for delay compensation by \eqref{EQN17} \\
    
    \vspace{0.1cm}
    // Compute relevance \eqref{EQ14}: $\mathrm{R(t)}=\bm{\varrho}_j-\bm{\varrho}_j^{\textbf{R}}$ \\
  }{// No reconstruction: $\mathrm{R(t)=0}$ \\
  }
  }{// Compute freshness \eqref{EQ14}: $\mathrm{F(t)}=\mathrm{t}-\mathrm{S(t)}$ \\
  }
}
\end{algorithm} 

  \item \textit{Semantic prediction policy:} The downsampling process generates two downsampled signals, namely $\mathrm{\varrho}_j^\mathrm{dD}\mathrm{(t)}$ and $\mathrm{\varrho}_j^\mathrm{qD}\mathrm{(t)}$. These downsampled signals are then compared with the local control inputs obtained from the SC, denoted as $\mathrm{\zeta}_j^\mathrm{p}\mathrm{(t)}$ and $\mathrm{\zeta}_j^\mathrm{q}\mathrm{(t)}$. Following the comparison, the semantic prediction policy comes into play. It reconstructs the signals necessary for attack compensation, denoted as $\bm{\varrho}_j(\mathrm{t}_a)=[{\mathrm{\varrho}_j^\mathrm{p}(\mathrm{t}_a) \hspace{0.2cm} \mathrm{\varrho}_j^\mathrm{q}(\mathrm{t}_a)}]^{\mathrm{T}}$, based on:
\vspace{-0.1cm}
\begin{equation}
\label{EQN14}
    \bm{\varrho}_j(\mathrm{t}_a)= [{\mathrm{\varrho}_j^\mathrm{dD}(\mathrm{t}_a) \hspace{0.2cm} \mathrm{\varrho}_j^\mathrm{qD}(\mathrm{t}_a)}]-\bm{\zeta}_j
\end{equation}

In SC, integrator accumulates error from the latest data. Persistent error accumulation due to cyber attacks may deviate the control system, causing undesired behaviors and potential instability. The generated reconstructed signals (guided by \eqref{EQN14}) act as feedback to local SCs, aiding in compensating for cyber attack impacts. In this regard, error signal is introduced into the prediction policy stage which is mathematically expressed as:
\vspace{-0.1cm}
\begin{equation}
    \label{EQN15}
    ||\bm{\varrho}_j(\mathrm{t}_a)||>\beta||e^{-\mathrm{t/T_c}}.[\mathrm{\varrho}_j^{\mathrm{dVC}} \hspace{0.2cm} \mathrm{\varrho}_j^{\mathrm{qVC}}]||
\end{equation}
Here, $\beta$ is a tunable value, and $\mathrm{T_c=K_p/K_i}$ is controller time constant of PI control loops $\mathrm{H_1(s)}$ and $\mathrm{H_2(s)}$. When the condition stated in \eqref{EQN15} is satisfied, triggers are generated. These triggers play a crucial role in the reconstruction of $\bm{\varrho}_j^{\mathrm{R}}(\mathrm{t}_a)$ using a sample-and-hold circuitry. Here, $\mathrm{t}_a$ denotes the instant of triggering. After the triggering moment, the assessment of semantic attributes ensues, encompassing parameters such as, freshness ($\mathrm{F}\mathrm{(t)}$), priority and relevance ($\mathrm{R(t)}$), delineated as:
\vspace{-0.1cm}
\begin{equation}
\label{EQ14}
    \mathrm{F}\mathrm{(t)}=\mathrm{t}-\mathrm{S(t)} \hspace{0.2cm}, \hspace{0.2cm} \mathrm{R(t)}=\bm{\varrho}_j-\bm{\varrho}_j^{\textbf{R}}
\end{equation}
Here, $\mathrm{S(t)}$ denotes the timestamp of the most recent packet successfully received at the destination by the time $\mathrm{t}$. Relevance computes the reconstruction error. Additionally, the inner control loop dynamics are prioritized to generate the reconstruction signals.\\

\item \textit{Feedback generation:} The reconstructed signals obtained from the previous step are then fed back to the SC module, incorporating tunable gains denoted as $\mathrm{g_1}$ and $\mathrm{g_2}$. These gains represent the adjustable parameters (here, $\mathrm{g_1}$ = 0.3 and $\mathrm{g_2}$ = 0.5) controlling the feedback mechanism, expressed as:
\vspace{-0.1cm}
\begin{equation}
\label{EQN16}
    \mathrm{\varrho}_{j}^{\mathrm{p}\varphi}(\mathrm{t}_a)=\mathrm{g_1}\mathrm{\varrho}_j^{\mathrm{Rp}}(\mathrm{t}_a) \hspace{0.2cm}, \hspace{0.2cm} \mathrm{\varrho}_{j}^{\mathrm{q}\varphi}(\mathrm{t}_a)=\mathrm{g_2}\mathrm{\varrho}_j^{\mathrm{Rq}}(\mathrm{t}_a)
\end{equation}
In this equation, $\bm{\varrho}_j^{\textbf{R}}(\mathrm{t}_a)=[\mathrm{\varrho}_j^{\mathrm{Rp}}(\mathrm{t}_a) \hspace{0.2cm} \mathrm{\varrho}_j^{\mathrm{Rq}}(\mathrm{t}_a)]^{\mathrm{T}}$ represents the reconstructed signals obtained at the triggering instant $\mathrm{t}_a$. These reconstructed signals are then added to the control inputs of the SC module as follows:
\vspace{-0.1cm}
\begin{equation}
\label{EQN17}   \mathrm{\zeta}_j^{\mathrm{pf}}\mathrm{(t)}=\mathrm{\zeta}_j^\mathrm{p}\mathrm{(t)}+\mathrm{\varrho}_{j}^{\mathrm{p}\varphi}(\mathrm{t}_a) \hspace{0.2cm}, \hspace{0.2cm} \mathrm{\zeta}_j^{\mathrm{qf}}\mathrm{(t)}=\mathrm{\zeta}_j^\mathrm{q}\mathrm{(t)}+\mathrm{\varrho}_{j}^{\mathrm{q}\varphi}(\mathrm{t}_a)
\end{equation}
Here, $\mathrm{\zeta}_j^{\mathrm{pf}}$ and $\mathrm{\zeta}_j^{\mathrm{qf}}$ are the final predictive inputs to the SC module. These inputs compensate for the cyber attacks experienced in the system. 
\end{enumerate}

\begin{figure}[h!]
        \centering
	    \includegraphics[clip, trim=0.5cm 3.9cm 3.9cm 0.7cm, width=0.9\linewidth]{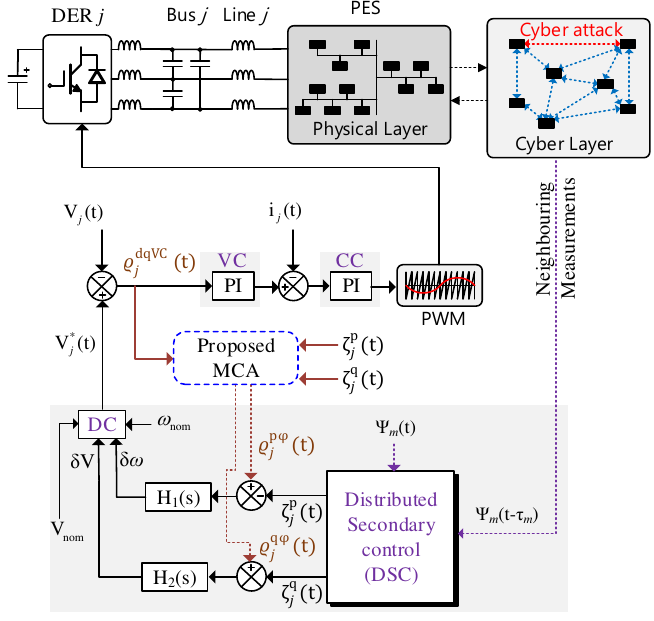}
	    \caption{Detailed control block diagram of the proposed MCA scheme.}
	    \label{fig:Control}
\end{figure}

A comprehensive visual representation of the equations \cref{EQN12,EQN13,EQN14,EQN15,EQ14,EQN16,EQN17}, involved in the proposed MCA is illustrated in the detailed control block diagram depicted in Fig. \ref{fig:Control}.

Further, a comprehensive exploration of the convergence analysis through a detailed mathematical analysis can be referred in \cite{A2}. Hence, the MCA aims to optimize information gathering, dissemination, and decision-making policies within cooperative networks, leading to a collectively optimal performance.

\section{Performance Evaluation}
\begin{figure}[h!]
        \centering
	    \includegraphics[ width=0.9\linewidth]{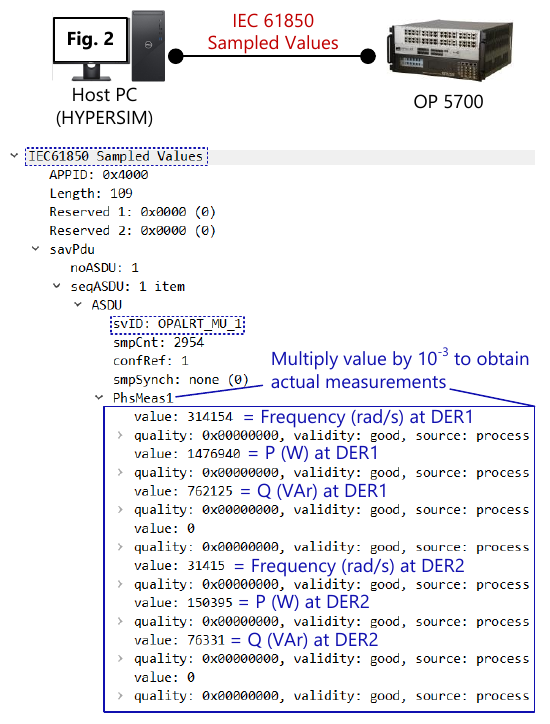}
	    \caption{Real-time simulation testbed with snapshot captured from wireshark, showing packet communicated through IEC 61850 sampled valued protocol.}
	    \label{fig:Testbed}
\end{figure}

The real-time simulation testbed \cite{ref19} is presented in Fig. \ref{fig:Testbed}, to validate the effectiveness and feasibility of the proposed MCA scheme. The setup involves an OP-5700 real-time simulator integrated with HYPERSIM software on the host PC, enabling modeling of the required test system. Efficient communication is achieved through seamless interconnection of the PC and OP-5700 via the IEC 61850 sampled values protocol. The details regarding network parameters and loads for the two test systems (i.e, the modified IEEE 69-bus system with nine DERs and real-world SCE 47-bus network with five DERs) can be obtained from \cite{ref16} and \cite{ref17}, respectively. The design and control parameters of the deployed DERs in these test systems are mentioned in Table \ref{tab:Parameters} and  \ref{tab:Parameters2}, respectively.  
\vspace{-0.5cm}
\begin{table}[h!]
  \centering
  \caption{Test System Parameters for IEEE 69-Bus Distribution System}
    \begin{tabular}{|c|c|c|}
    \hline
    \multicolumn{3}{|c|}{\textbf{Parameters for DER}} \\
    \hline
    Parameter & Symbol & Rating \\
    \hline
    Active power droop coefficient & $\mathrm{m^{p}}$ & 9.4$\times$10$^{-5}$ rad/(W.s) \\
    Reactive power droop coefficient & $\mathrm{n^{q}}$ & 1.3$\times$10$^{-3}$V/VAr \\
    Proportional gain (CC, VC) & $\mathrm{K_{p}^{i}}$, $\mathrm{K_{p}^{V}}$ & 0.2, 50 \\
    \rule{0pt}{3ex}
    Integral gain (CC, VC) & $\mathrm{K_{i}^{i}}$, $\mathrm{K_{i}^{V}}$ & 1, 100 \\
    \hline
    \multicolumn{3}{|c|}{\textbf{Secondary control (SC) parameters}} \\
    \hline
    \rule{0pt}{3ex} 
    Proportional gain & $\mathrm{K_{p}^{S\omega}}$, $\mathrm{K_{p}^{SV}}$ & 0.1, 0.1 \\
    \rule{0pt}{3ex}  
    Integral gain & $\mathrm{K_{i}^{S\omega}}$, $\mathrm{K_{i}^{SV}}$ & 100, 10 \\
    \hline
    \end{tabular}%
  \label{tab:Parameters}%
\end{table} 
\begin{table}[h!]
  \centering
  \caption{Test System Parameters for Real World SCE 47-Bus Network}
    \begin{tabular}{|c|c|c|}
    \hline
    \multicolumn{3}{|c|}{\textbf{Parameters for DER}} \\
    \hline
    Parameter & Symbol & Rating \\
    \hline
    Active power droop coefficient & $\mathrm{m^{p}}$ & 9.4$\times$10$^{-5}$ rad/(W.s) \\
    Reactive power droop coefficient & $\mathrm{n^{q}}$ & 1.3$\times$10$^{-3}$V/VAr \\
    Proportional gain (CC, VC) & $\mathrm{K_{p}^{i}}$, $\mathrm{K_{p}^{V}}$ & 0.1, 40 \\
    \rule{0pt}{3ex}
    Integral gain (CC, VC) & $\mathrm{K_{i}^{i}}$, $\mathrm{K_{i}^{V}}$ & 0.5, 80 \\
    \hline
    \multicolumn{3}{|c|}{\textbf{Secondary control (SC) parameters}} \\
    \hline
    \rule{0pt}{3ex} 
    Proportional gain & $\mathrm{K_{p}^{S\omega}}$, $\mathrm{K_{p}^{SV}}$ & 0.1, 0.1 \\
    \rule{0pt}{3ex}  
    Integral gain & $\mathrm{K_{i}^{S\omega}}$, $\mathrm{K_{i}^{SV}}$ & 10, 1.5 \\
    \hline
    \end{tabular}%
  \label{tab:Parameters2}%
\end{table}

The performance of proposed MCA scheme is rigorously evaluated in contrast to a conventional DSC scheme, referred to as `without MCA' in subsequent sections. The assessment includes various scenarios, such as latency attacks, data dropouts, TSAs, and FDIAs. The subsequent sections detail the results of these evaluations.

\subsection{Performance on the Modified IEEE 69-Bus Distribution System}
\subsubsection{System under latency attack}
\begin{figure}[h!]
        \centering
	    \includegraphics[clip, trim=0.5cm 6cm 8cm 0.7cm, width=1\linewidth]{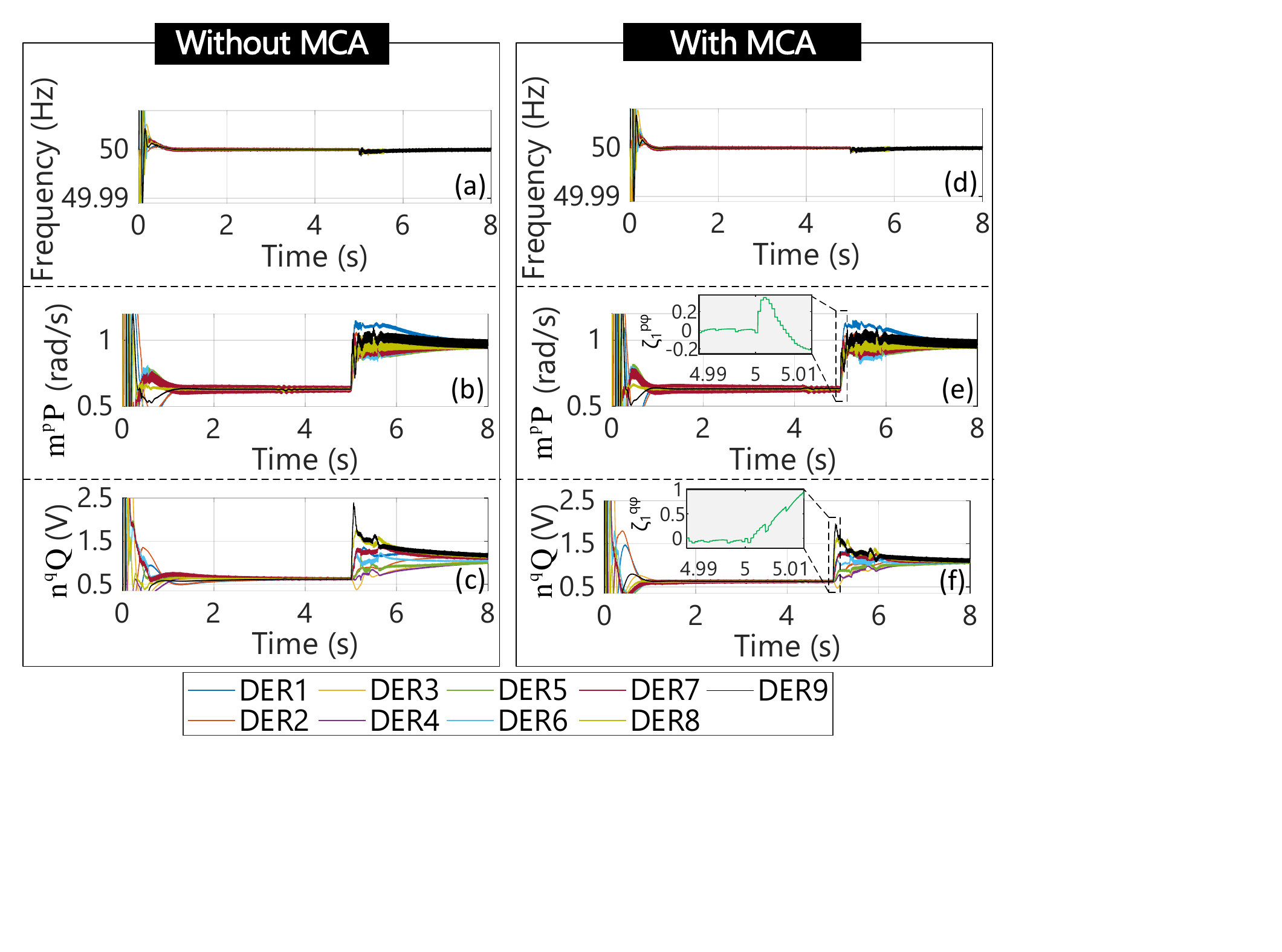}
	    \caption{Performance during latency attack ($\tau_m$=0.05 s) for \textbf{(a)} frequency; \textbf{(b)} active power sharing; \textbf{(c)} reactive power sharing, without MCA are shown. Further, the time-domain signals for \textbf{(d)} frequency; \textbf{(e)} active power sharing; \textbf{(f)} reactive power sharing, with MCA are shown.}
	    \label{fig:LA_69}
\end{figure}

A latency attack was initiated on the system with a time delay of $\tau_m$ = 0.05 s, followed by a load variation at 5 s. While the system remained stable, the convergence time for achieving SC objectives was longer without the proposed scheme (Fig. \ref{fig:LA_69}(a), (b), and (c))  compared to the proposed scheme (Fig. \ref{fig:LA_69}(d), (e), and (f)) due to the resulting reconstructed signals ($\mathrm{\zeta}_{j}^{\mathrm{p}\varphi}$ and $\mathrm{\zeta}_{j}^{\mathrm{q}\varphi}$). These reconstructed signals generated through semantic sampling, actively synchronizes the error signals at primary and secondary controllers to compensate for the latency attack.

\begin{figure}[h!]
        \centering
	    \includegraphics[clip, trim=0.5cm 3.5cm 7cm 0.7cm, width=1\linewidth]{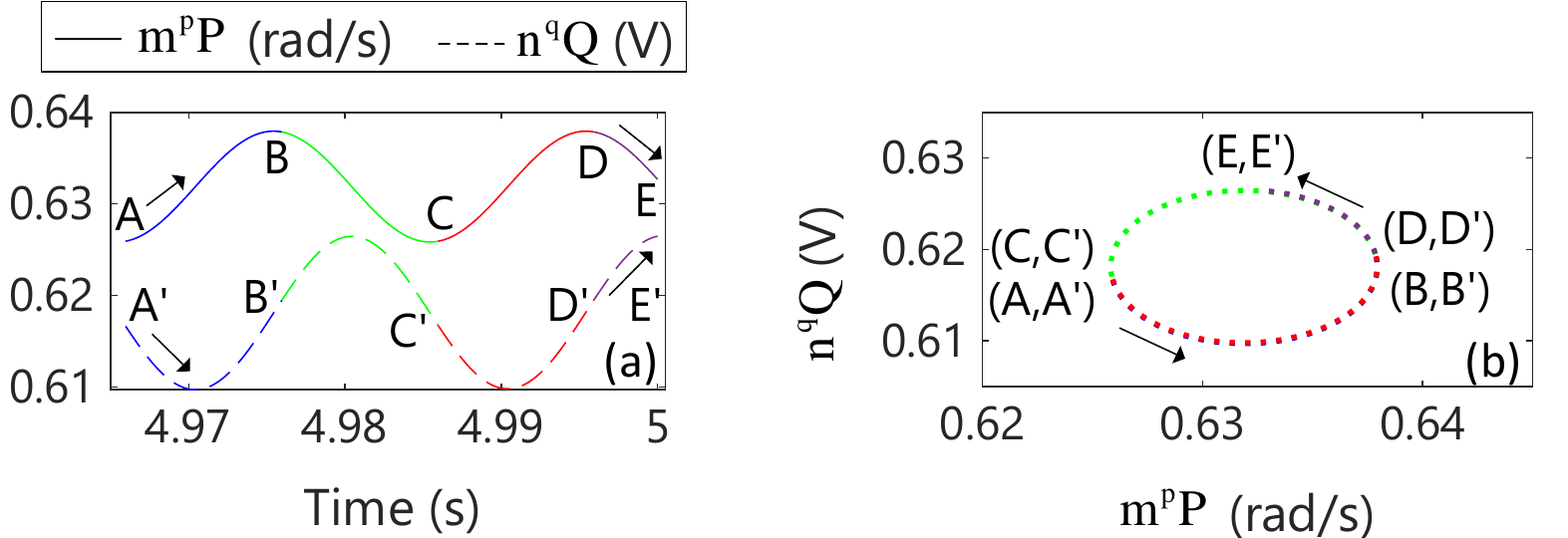}
	    \caption{\textbf{(a)} The time-domain plots of active and reactive power sharing are shown by solid and dashed lines respectively, before cyber attack; and \textbf{(b)} phase portrait of a DER in $\mathrm{n}^{\mathrm{q}}\mathrm{Q}$-$\mathrm{m}^{\mathrm{p}}\mathrm{P}$ domain before cyber attack is shown.}
	    \label{fig:Before}
\end{figure}

The variations of active and reactive power sharing are depicted in Fig. \ref{fig:Before}(a), with time on the $\mathrm{X}$-axis and power on the $\mathrm{Y}$-axis. The corresponding $\mathrm{Y}$-axis coordinates (A, B, C, D, E) and (A', B', C', D', E') for active and reactive power sharing curves are plotted against each other ($\mathrm{n}^{\mathrm{q}}\mathrm{Q}$-$\mathrm{m}^{\mathrm{p}}\mathrm{P}$ as $\mathrm{Y}$-$\mathrm{X}$) to generate a phase portrait, as illustrated in Fig. \ref{fig:Before}(b). These phase portraits serve as a basis for result evaluation.

\begin{figure}[h!]
        \centering
	    \includegraphics[clip, trim=0.5cm 3cm 7.5cm 0.7cm, width=1\linewidth]{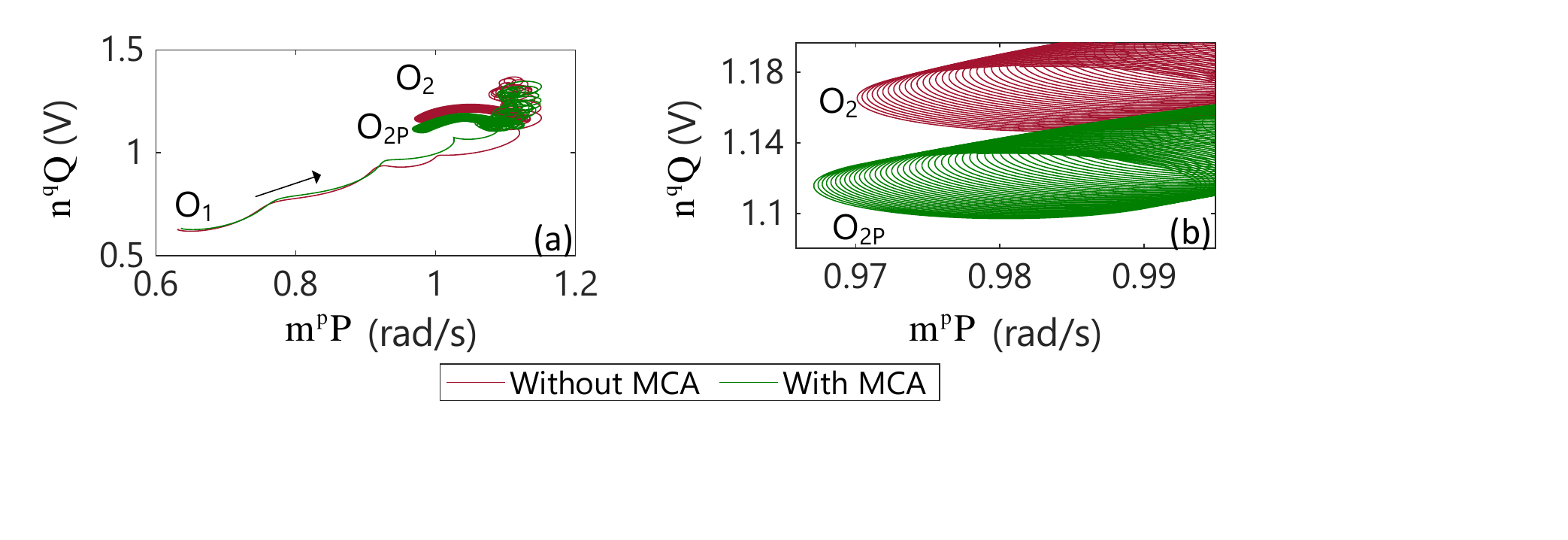}
	    \caption{\textbf{(a)} Phase portrait of a DER 1 in $\mathrm{n}^{\mathrm{q}}\mathrm{Q}$-$\mathrm{m}^{\mathrm{p}}\mathrm{P}$ domain during latency attack ($\tau_m$=0.05 s) is shown. The operating point $\mathrm{O}_{\mathrm{1}}$ (before cyber attack) traverses to $\mathrm{O}_{\mathrm{2}}$ and $\mathrm{O}_{\mathrm{2P}}$, representing operating points obtained without and with MCA respectively. \textbf{(b)} Zoomed plot for (a) is shown.}
	    \label{fig:A_LA}
\end{figure}

In Fig. \ref{fig:LA_69}, the MCA scheme notably accelerates the convergence time for SC objectives. Similar trends are observed in the phase portrait, as depicted in Fig. \ref{fig:A_LA}(a), which is further magnified in Fig. \ref{fig:A_LA}(b). The shift of operating point $\mathrm{O}_{\mathrm{1}}$ (prior to the cyber attack) to $\mathrm{O}_{\mathrm{2}}$ and $\mathrm{O}_{\mathrm{2P}}$ signifies the operating points achieved without and with MCA, respectively.

\subsubsection{System under latency attack and data dropout}
\begin{figure}[h!]
        \centering
	    \includegraphics[clip, trim=0.5cm 3cm 7.5cm 0.7cm, width=1\linewidth]{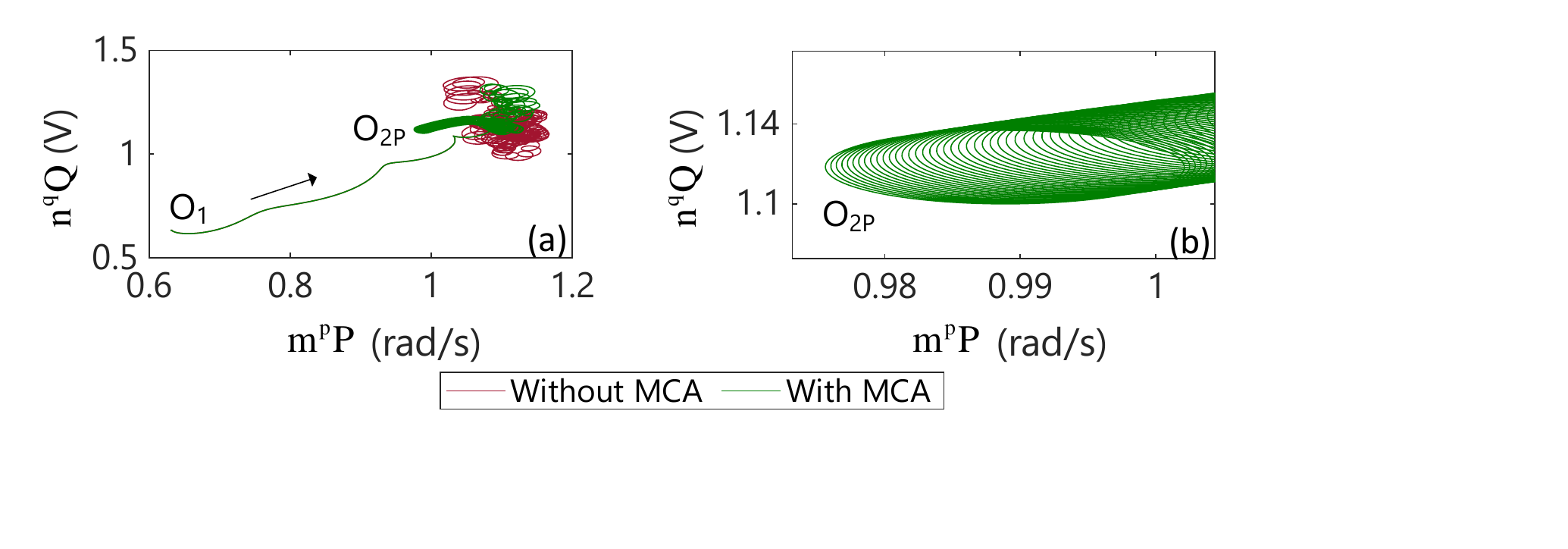}
	    \caption{\textbf{(a)} Phase portrait of a DER 1 in $\mathrm{n}^{\mathrm{q}}\mathrm{Q}$-$\mathrm{m}^{\mathrm{p}}\mathrm{P}$ domain during latency attack ($\tau_m$=0.05 s) and 10\% data dropout is shown. The operating point $\mathrm{O}_{\mathrm{1}}$ (before cyber attack) traverses to $\mathrm{O}_{\mathrm{2}}$ and $\mathrm{O}_{\mathrm{2P}}$, representing operating points obtained without and with MCA respectively. \textbf{(b)} Zoomed plot for (a) is shown.}
	    \label{fig:A_LA_DD}
\end{figure}

For a latency attack ($\tau_m$ = 0.05 s) combined with 10\% data dropout and load variation at t = 5 s, Fig. \ref{fig:A_LA_DD}(a)  indicates that the system without the MCA scheme fails to reach a steady-state operating point (even within 3 s of attack initiation), unlike the system with MCA. This is attributed to the signal reconstruction from the local controller steering the control process during the attack. The Fig. \ref{fig:A_LA_DD}(b) is zoomed plot for Fig. \ref{fig:A_LA_DD}(a). Additionally, it is noted that the convergence time is extended in this scenario for the system without MCA, compared to the case of the sole latency attack.

\subsubsection{System under TSA}
\begin{figure}[h!]
        \centering
	    \includegraphics[clip, trim=0.5cm 3cm 7.5cm 0.7cm, width=1\linewidth]{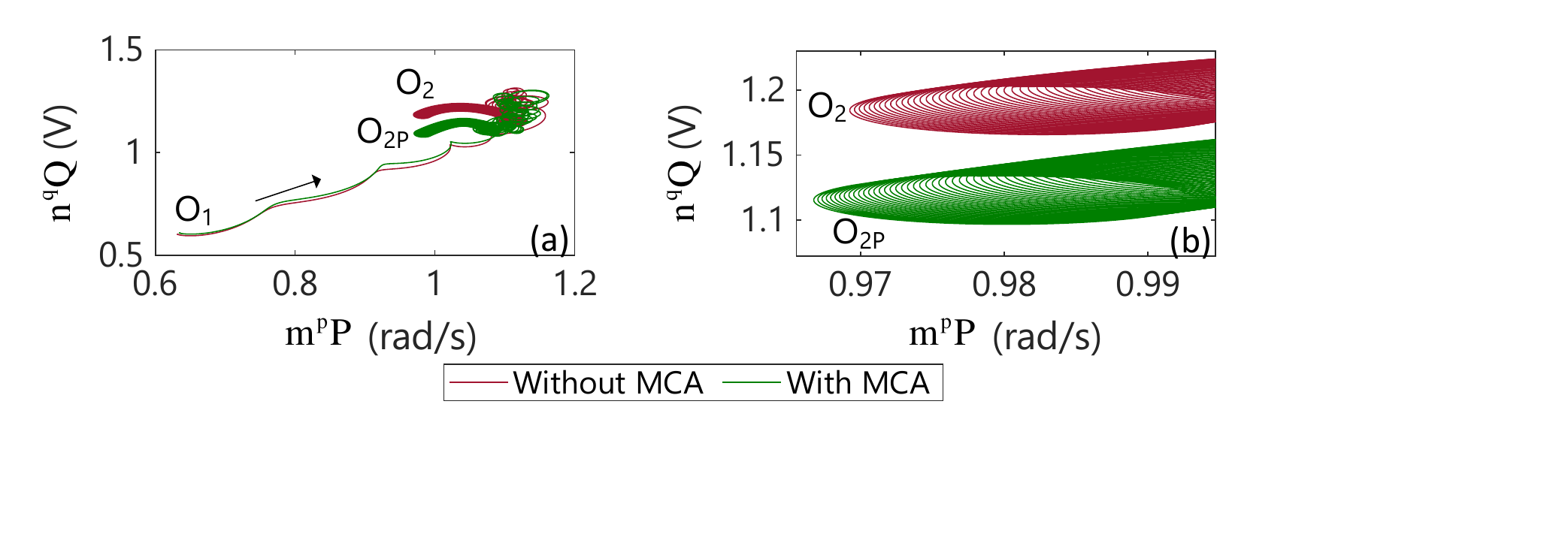}
	    \caption{\textbf{(a)} Phase portrait of a DER 1 in $\mathrm{n}^{\mathrm{q}}\mathrm{Q}$-$\mathrm{m}^{\mathrm{p}}\mathrm{P}$ domain during TSA is shown. The operating point $\mathrm{O}_{\mathrm{1}}$ (before cyber attack) traverses to $\mathrm{O}_{\mathrm{2}}$ and $\mathrm{O}_{\mathrm{2P}}$, representing operating points obtained without and with MCA respectively. \textbf{(b)} Zoomed plot for (a) is shown.}
	    \label{fig:A_TSA}
\end{figure}

Likewise, the phase portrait for the system subjected to a TSA attack (with load variation at 5 s) is depicted in Fig. \ref{fig:A_TSA}(a).  Evidently, the proposed scheme leads to enhanced dynamic performance due to signal reconstruction. The zoomed-in plot in Fig. \ref{fig:A_TSA}(b) further demonstrates that the MCA scheme reduces the convergence time for achieving SC objectives. 

\subsubsection{System under balanced FDIA}
\begin{figure}[h!]
        \centering
	    \includegraphics[clip, trim=0.5cm 3cm 8cm 0.7cm, width=1\linewidth]{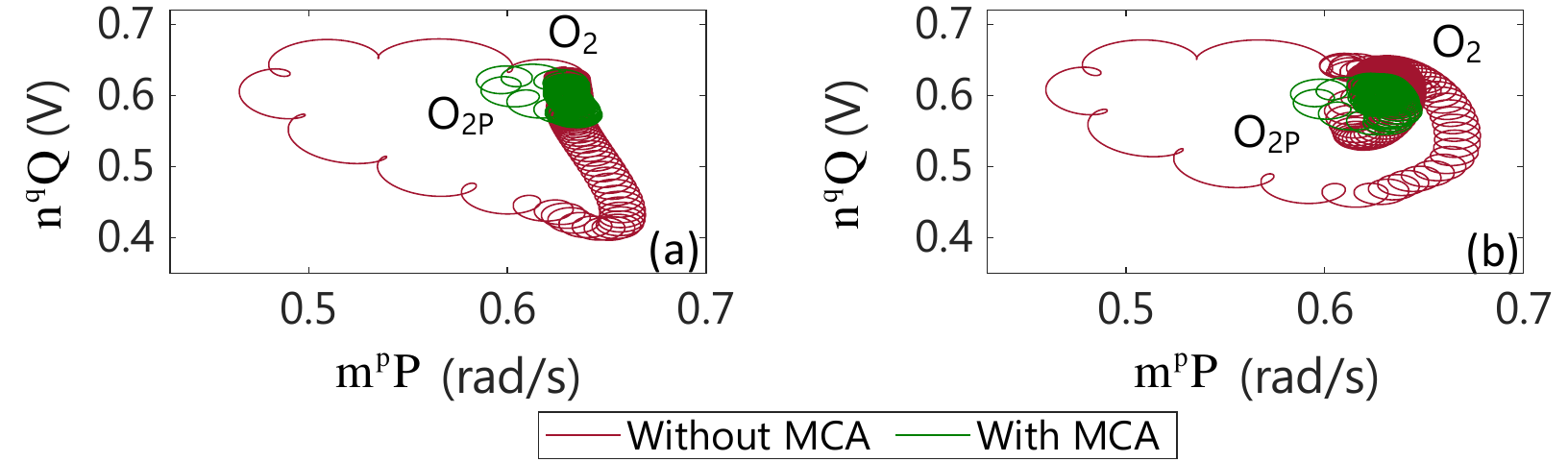}
	    \caption{Phase portrait of a DER 1 in $\mathrm{n}^{\mathrm{q}}\mathrm{Q}$-$\mathrm{m}^{\mathrm{p}}\mathrm{P}$ domain during \textbf{(a)} balanced FDIA; and \textbf{(b)} balanced FDIA with latency attack ($\tau_m$=0.05 s) is shown. $\mathrm{O}_{\mathrm{2}}$ and $\mathrm{O}_{\mathrm{2P}}$ represents operating points obtained without and with MCA respectively.}
	    \label{fig:A_FDIAandLA}
\end{figure}

The phase portrait for the system under a balanced FDIA at 5 s is presented in Fig. \ref{fig:A_FDIAandLA}(a). Additionally, Fig. \ref{fig:A_FDIAandLA}(b) displays the phase portrait for the system exposed to above case along with a time delay of $\tau_m$ = 0.05 s. Clearly, the proposed scheme demonstrates improved dynamic performance through signal reconstruction aided through risk-aware semantic sampling in both scenarios. The semantic sampling expedites convergence through the transmission of significant information.

\subsubsection{System under system configuration variation and latency attack}
Consider a modified 69-bus distribution system with nine DERs, as shown in Fig. \ref{fig:69bus}. Let this system is initially in network topology N1, with all tie-line switches being open. Suppose the system rearranges at 5 s to a new network topology. This is denoted as N2, \begin{figure}[h!]
        \centering
	    \includegraphics[clip, trim=0.5cm 3cm 7.5cm 0.7cm, width=1\linewidth]{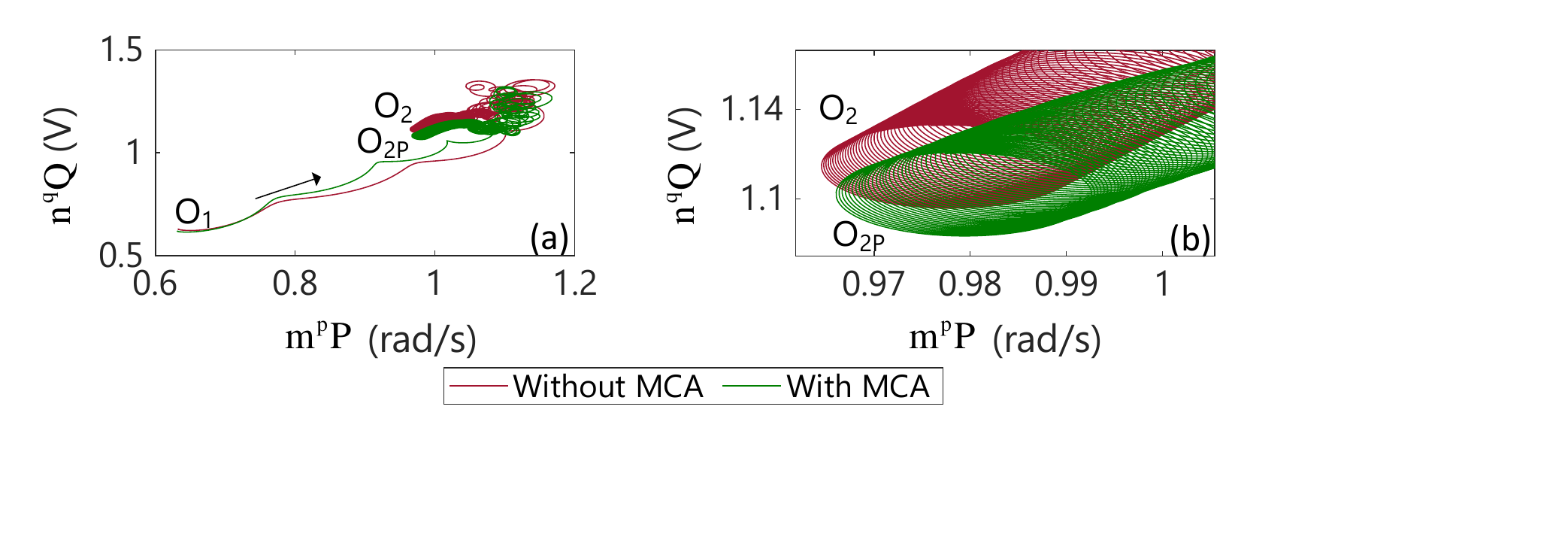}
	    \caption{\textbf{(a)} Phase portrait of a DER 1 in $\mathrm{n}^{\mathrm{q}}\mathrm{Q}$-$\mathrm{m}^{\mathrm{p}}\mathrm{P}$ domain during latency attack ($\tau_m$=0.05 s) with system reconfiguration is shown. The operating point $\mathrm{O}_{\mathrm{1}}$ (before cyber attack) traverses to $\mathrm{O}_{\mathrm{2}}$ and $\mathrm{O}_{\mathrm{2P}}$, representing operating points obtained without and with MCA respectively. \textbf{(b)} Zoomed plot for (a) is shown.}
	    \label{fig:A_Tie_LA}
\end{figure} with all tie-line switches being closed. This is further accompanied by a latency attack with time delay of $\tau_m$ = 0.05 s. As evident from Fig. \ref{fig:A_Tie_LA}(a), the steady state operating point is attained quickly as compared to the absence of the MCA scheme. Fig. \ref{fig:A_Tie_LA}(b) presents a magnified view of Fig. \ref{fig:A_Tie_LA}(a). This improved convergence can be attributed to the local reconstruction error provided by the SC, enabled by semantic sampling approach in the proposed scheme.

\subsection{Performance on the real-world SCE 47-bus network}
\subsubsection{System under balanced FDIA}
\begin{figure}[h!]
        \centering
	    \includegraphics[clip, trim=0.5cm 5.8cm 8cm 0.7cm, width=1\linewidth]{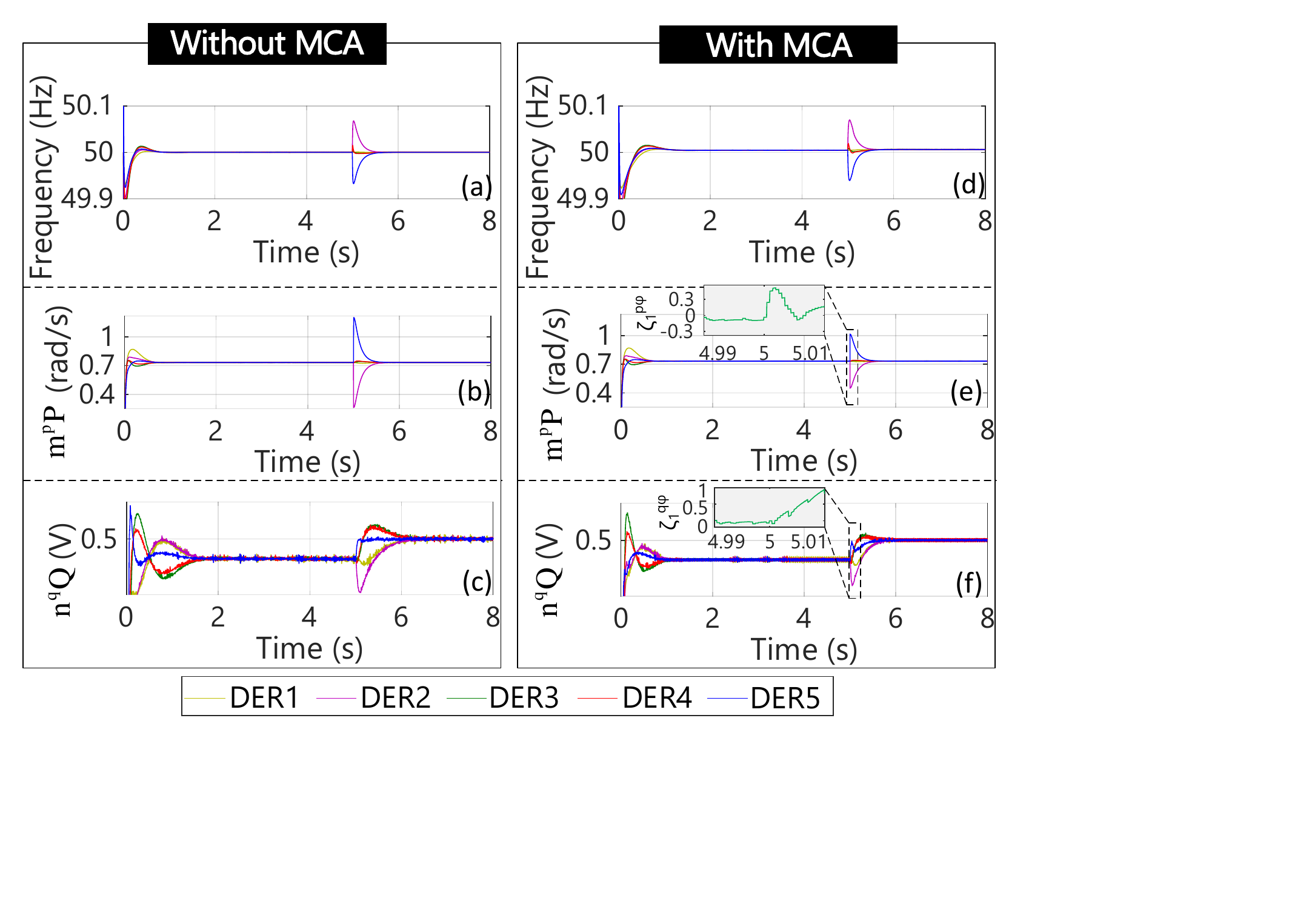}
	    \caption{The time-domain signals during balanced FDIA for \textbf{(a)} frequency; \textbf{(b)} active power sharing; \textbf{(c)} reactive power sharing, without MCA are shown. Further, the time-domain signals for \textbf{(d)} frequency; \textbf{(e)} active power sharing; \textbf{(f)} reactive power sharing, with MCA are shown.}
	    \label{fig:FDIA_S_47}
\end{figure}

A balanced FDIA was initiated on the system at t = 5 s. Despite system stability being maintained, the convergence time for achieving SC objectives was extended in the absence of the proposed scheme (Fig. \ref{fig:FDIA_S_47}(a), (b), and (c)) when contrasted with its presence (Fig. \ref{fig:FDIA_S_47}(d), (e), and (f)). The reconstructed signals ($\mathrm{\zeta}_{j}^{\mathrm{p}\varphi}$ and $\mathrm{\zeta}_{j}^{\mathrm{q}\varphi}$), generated through the MCA scheme improves the dynamic performance of the system under FDIA. Consequently, the convergence time is reduced.

\subsubsection{System under unbalanced FDIA}
\begin{figure}[h!]
        \centering
	    \includegraphics[clip, trim=0.5cm 3cm 8cm 0.7cm, width=1\linewidth]{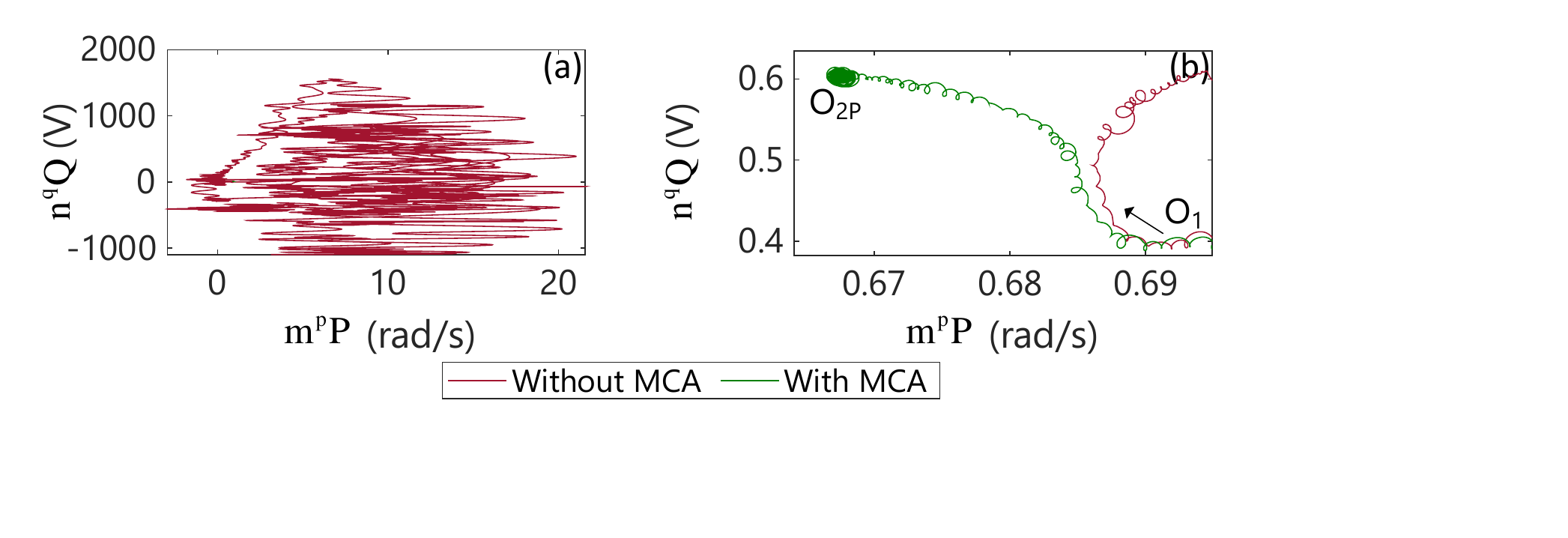}
	    \caption{\textbf{(a)} Phase portrait of a DER 1 in $\mathrm{n}^{\mathrm{q}}\mathrm{Q}$-$\mathrm{m}^{\mathrm{p}}\mathrm{P}$ domain during unbalanced FDIA is shown. \textbf{(b)} Zoomed plot for (a) is shown, where the operating point $\mathrm{O}_{\mathrm{1}}$ (before cyber attack) traverses to $\mathrm{O}_{\mathrm{2}}$ and $\mathrm{O}_{\mathrm{2P}}$, representing operating points obtained without and with MCA respectively.}
	    \label{fig:A_FDIA_U_47}
\end{figure}

The phase portrait of the system subjected to an unbalanced FDIA at t = 5 s is depicted in Fig. \ref{fig:A_FDIA_U_47}(a). System instability is evident without the MCA scheme, while the MCA scheme achieves a steady-state operating point. This achievement is attributed to signal reconstruction facilitated by risk-aware semantic sampling.

\begin{table*}[b!]
\caption{Comparative Analysis of the Proposed Monolithic Cybersecurity Architecture (MCA) for PES.}
\centering
\def\arraystretch{0.6}
\label{table:comp2}
\begin{tabular}{|>{\color{black}}p{0.04\linewidth}|>{\color{black}}p{0.2\linewidth}|>{\color{black}}p{0.15\linewidth}|>{\color{black}}p{0.15\linewidth}|>{\color{black}}p{0.15\linewidth}|>{\color{black}}p{0.15\linewidth}|}
\hline
S.No. & Features & \cite{RF23} & \cite{RF25} & \cite{RF26} & \textbf{Proposed scheme}\\
\hline
1. & Computational burden & Medium & Medium & High & Low\\
2. & Distributed concept & \xmark & \xmark & \xmark & \cmark\\
3. & Resilient to latency attacks & \cmark & \cmark & Not tested & \cmark \\
4. & Resilient to TSAs & Not tested & Not tested & Only detection & \cmark \\
5. & Resilient to data dropouts & Not tested & Not tested & Not tested & \cmark \\
6. & Resilient to FDIAs & Not tested & Not tested & Not tested & \cmark \\
7. & Robust to loading variations & \cmark & \cmark & \cmark & \cmark \\
8. & Model-agnostic & \xmark & \xmark & \cmark & \cmark \\
9. & Supports dynamic cyber graphs & \cmark & Not tested & Not tested & \cmark \\
10. & Scalable & \cmark & Not tested & Not tested & \cmark \\
\hline
\end{tabular}
\end{table*}

\subsubsection{System under balanced FDIA and latency attack}
\begin{figure}[h!]
        \centering
	    \includegraphics[clip, trim=0.5cm 3cm 8cm 0.7cm, width=1\linewidth]{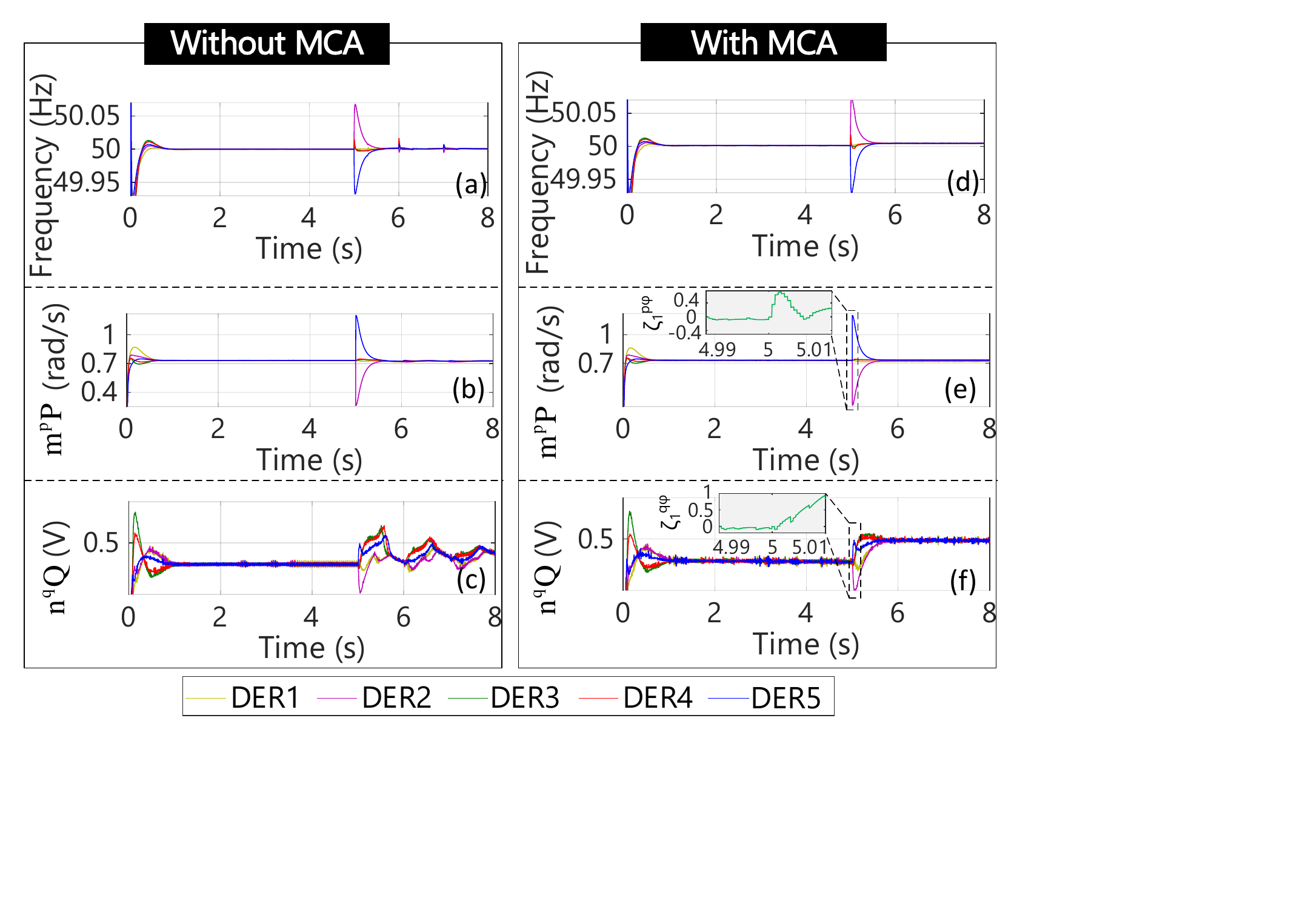}
	    \caption{The time-domain signals during balanced FDIA with latency attack ($\tau_m$=0.05 s) for \textbf{(a)} frequency; \textbf{(b)} active power sharing; \textbf{(c)} reactive power sharing, without MCA are shown. The time-domain signals for \textbf{(d)} frequency; \textbf{(e)} active power sharing; \textbf{(f)} reactive power sharing, with MCA are shown.}
	    \label{fig:FDIA_LA_47}
\end{figure}

A FDIA was initiated on the system with a time delay of $\tau_m$ = 0.05 s at 5 s. While system stability was maintained, achieving SC objectives took longer without the proposed scheme (Fig. \ref{fig:FDIA_LA_47}(a), (b), and (c)) compared to with the proposed scheme (Fig. \ref{fig:FDIA_LA_47}(d), (e), and (f)). Additionally, the convergence time is extended in this scenario (without MCA) compared to the case of balanced FDIA attack alone. The proposed scheme actively captures the semantic attributes to enhance its dynamic behavior and reconstruct the required signal to compensate for both FDIA and latency attack at the same time.

\subsubsection{System under cyber graph variations and latency attack}
\begin{figure}[h!]
        \centering
	    \includegraphics[clip, trim=0.5cm 3cm 8cm 0.7cm, width=1\linewidth]{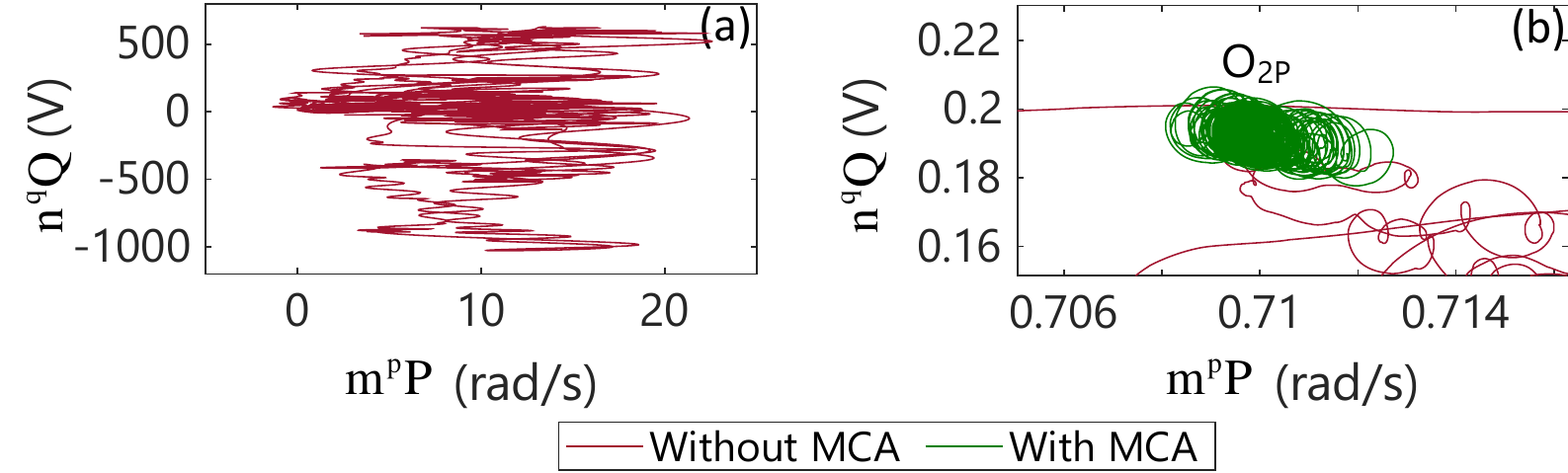}
	    \caption{\textbf{(a)} Phase portrait of a DER 1 in $\mathrm{n}^{\mathrm{q}}\mathrm{Q}$-$\mathrm{m}^{\mathrm{p}}\mathrm{P}$ domain during latency attack ($\tau_m$=0.05 s) with cyber topology variation from T1 to T2 is shown. \textbf{(b)} Zoomed plot for (a) is shown, where $\mathrm{O}_{\mathrm{2P}}$ is the operating point with MCA.}
	    \label{fig:A_Top_47}
\end{figure}

The system was assessed for cyber graphs variation from fully-connected (T1) to spanning tree (T2), as shown in Fig. \ref{fig:Map}(c). This is accompanied with a time delay of $\tau_m$ = 0.05 s. Fig. \ref{fig:A_Top_47}(a) illustrates that the system trends towards instability due to the sparse network and additional signal delay. The agents couldn't update their states continually, thereby slowing convergence. However, the risk-aware semantic sampling synchronizes error signals at primary and secondary controllers to generate reconstructed signals, rendering the proposed scheme robust to dynamic cyber graph variations and latency attacks (Fig. \ref{fig:A_Top_47}(b)).

A comparative evaluation of our proposed MCA in contrast to established methodologies is presented in Table \ref{table:comp2}. The proposed methodology in this study distinguishes itself as a computationally efficient solution by incorporating distributed concepts that enhance resilience against data availability and data integrity attacks into a monolithic plane. Its model-agnostic nature further facilitates an easy and streamlined implementation. Additionally, its support for dynamic cyber graphs emphasizes its practicality and flexibility. Furthermore, the scalability of the proposed scheme has been rigorously tested and validated. Consequently from an implementation perspective, the semantic edge intelligence presented by our approach emerges as a highly promising and commercially viable solution to address the intricate challenges within the realm of PES.  

\section{Conclusions and Future Scope of Work}
This paper introduces methodology to tackle challenges posed by random communication delays, data dropouts, and data integrity attacks single-handedly using a monolithic cybersecurity architecture in PES, incorporating semantic intelligence. The proposed MCA framework leverages the inner control loop dynamics of inverter-integrated DERs and incorporates key semantic attributes, including Priority, Freshness, and Relevance to generate localized compensation signals. This approach has the potential to revolutionize the cybersecurity landscape of PES by simultaneously addressing multiple types of attacks, offering a one-stop solution. Notably, it eliminates the requirement for various intricate models and their complex deployment, a common practice in existing methodologies. Real-time simulations conducted in an OPAL-RT environment validate the effectiveness of this approach.

In forthcoming research, demand response scenarios will be investigated through a semantic framework, with a focus on flexibility estimation and synchronizing market requests and bid parameters. The findings will contribute to the development of an innovative demand response model suitable for systems with diverse resources and varied communication protocols.Additionally, semantic-based quantum communication architecture will be explored to expedite fault detection and localization within PES. This enhancement will lead to reduced response times and minimized downtime during substantial disturbances, thereby strengthening system resiliency.

\ifCLASSOPTIONcaptionsoff
  \newpage
\fi



\begin{thebibliography}{1}

\bibitem{ref1}
A. Singhal, T. L. Vu and W. Du, ``Consensus Control for Coordinating Grid-Forming and Grid-Following Inverters in Microgrids," \textit{IEEE Trans. Smart Grid}, vol. 13, no. 5, pp. 4123-4133, Sept. 2022, doi: 10.1109/TSG.2022.3158254.

\bibitem{ref2}
S. Patel, S. Chakraborty, B. Lundstrom, S. M. Salapaka and M. V. Salapaka, ``Isochronous Architecture-Based Voltage-Active Power Droop for Multi-Inverter Systems," \textit{IEEE Trans. Smart Grid}, vol. 12, no. 2, pp. 1088-1103, March 2021, doi: 10.1109/TSG.2020.3037159.

\bibitem{ref3}
D. C. Mazur, R. D. Quint and V. A. Centeno, ``Time Synchronization of Automation Controllers for Power Applications," \textit{IEEE Trans. Ind. Appl.}, vol. 50, no. 1, pp. 25-32, Jan.-Feb. 2014, doi: 10.1109/TIA.2013.2267710.

\bibitem{ref4}
Y. Han, K. Zhang, H. Li, E. A. A. Coelho and J. M. Guerrero, ``MAS-Based Distributed Coordinated Control and Optimization in Microgrid and Microgrid Clusters: A Comprehensive Overview," \textit{IEEE Trans. Power Electron.}, vol. 33, no. 8, pp. 6488-6508, Aug. 2018, doi: 10.1109/TPEL.2017.2761438.

\bibitem{ref5}
K. Gupta, S. Sahoo, R. Mohanty, B. Ketan Panigrahi and F. Blaabjerg, ``Distinguishing Between Cyber Attacks and Faults in Power Electronic Systems—A Noninvasive Approach," \textit{IEEE J. Emerg. Sel. Topics Power Electron.}, vol. 11, no. 2, pp. 1578-1588, April 2023, doi: 10.1109/JESTPE.2022.3221867.

\bibitem{ref6}
S. Sahoo, F. Blaabjerg, and T. Dragicevic, \textit{Cyber Security for Microgrids}. IET, 2022, doi: 10.1049/PBPO196E.

\bibitem{ref7}
R. Kateb, P. Akaber, M. H. K. Tushar, A. Albarakati, M. Debbabi and C. Assi, ``Enhancing WAMS Communication Network Against Delay Attacks," \textit{IEEE Trans. Smart Grid}, vol. 10, no. 3, pp. 2738-2751, May 2019, doi: 10.1109/TSG.2018.2809958.

\bibitem{ref8}
K. Gupta, S. Sahoo, B. K. Panigrahi and C. Konstantinou, ``Impact Assessment of Data Integrity Attacks in MVDC Shipboard Power Systems," \textit{IEEE Electric Ship Technologies Symposium (ESTS)}, Alexandria, VA, USA, 2023, pp. 512-519, doi: 10.1109/ESTS56571.2023.10220547.

\bibitem{ref9}
L. Sheng, G. Lou, W. Gu, S. Lu, S. Ding and Z. Ye, ``Optimal Communication Network Design of Microgrids Considering Cyber-Attacks and Time-Delays," \textit{IEEE Trans. Smart Grid}, vol. 13, no. 5, pp. 3774-3785, Sept. 2022, doi: 10.1109/TSG.2022.3169343.

\bibitem{ref10}
M. Kumar, ``Resilient PIDA Control Design Based Frequency Regulation of Interconnected Time-Delayed Microgrid Under Cyber-Attacks," \textit{IEEE Trans. Ind. Appl.}, vol. 59, no. 1, pp. 492-502, Jan.-Feb. 2023, doi: 10.1109/TIA.2022.3205280.

\bibitem{ref11}
T. Yang, Y. He and G. -P. Liu, ``Distributed Voltage Restoration of AC Microgrids Under Communication Delays: A Predictive Control Perspective," \textit{IEEE Trans. Circuits Syst. I, Reg. Papers}, vol. 69, no. 6, pp. 2614-2624, June 2022, doi: 10.1109/TCSI.2022.3163204.

\bibitem{ref12}
A. Mohammad Saber, A. Youssef, D. Svetinovic, H. H. Zeineldin and E. F. El-Saadany, ``Anomaly-Based Detection of Cyberattacks on Line Current Differential Relays," \textit{IEEE Trans. Smart Grid}, vol. 13, no. 6, pp. 4787-4800, Nov. 2022, doi: 10.1109/TSG.2022.3185764.

\bibitem{ref12a}
J. Zhang et al., ``Machine Learning-Based Cyber-Attack Detection in Photovoltaic Farms," \textit{IEEE Open J. Power Electron.}, vol. 4, pp. 658-673, 2023, doi: 10.1109/OJPEL.2023.3309897.

\bibitem{ref13a}
M. Alizadeh, H. A. Abyaneh and K. Rouzbehi, ``An Enhanced Distributed Event-Triggered Mechanism of Cyber-Switched Communications for Control of Islanded AC Microgrids," \textit{IEEE Sys. J.}, vol. 17, no. 1, pp. 1501-1511, March 2023, doi: 10.1109/JSYST.2022.3206026.

\bibitem{ref13b}
H. Yang, C. Deng, X. Xie and L. Ding, ``Distributed Resilient Secondary Control for AC Microgrid Under FDI Attacks," \textit{IEEE Trans. Circuits Syst. II-Express Briefs}, vol. 70, no. 7, pp. 2570-2574, July 2023, doi: 10.1109/TCSII.2023.3245282.

\bibitem{B1}
H. Xie, Z. Qin, G. Y. Li and B. -H. Juang, ``Deep Learning Enabled Semantic Communication Systems," \textit{IEEE Trans. Signal Process.}, vol. 69, pp. 2663-2675, 2021, doi: 10.1109/TSP.2021.3071210.

\bibitem{B2}
M. Kountouris and N. Pappas, ``Semantics-Empowered Communication for Networked Intelligent Systems," \textit{IEEE Commun. Mag.}, vol. 59, no. 6, pp. 96-102, June 2021, doi: 10.1109/MCOM.001.2000604.

\bibitem{B3}
T. Han, Q. Yang, Z. Shi, S. He and Z. Zhang, ``Semantic-Preserved Communication System for Highly Efficient Speech Transmission," \textit{IEEE J. Sel. Areas Commun.}, vol. 41, no. 1, pp. 245-259, Jan. 2023, doi: 10.1109/JSAC.2022.3221952.

\bibitem{B4}
P. Popovski, O. Simeone, F. Boccardi, D. Gunduz, and O. Sahin,
``Semantic-effectiveness filtering and control for post-5G wireless connectivity,” \textit{J. Indian Inst. Sci.}, vol. 100, no. 2, pp. 435–443, Apr. 2020.

\bibitem{B5}
X. Luo, H. -H. Chen and Q. Guo, ``Semantic Communications: Overview, Open Issues, and Future Research Directions," \textit{IEEE Wireless Commun.}, vol. 29, no. 1, pp. 210-219, February 2022, doi: 10.1109/MWC.101.2100269.

\bibitem{ref14}
M. Kountouris and N. Pappas, ``Semantics-Empowered Communication for Networked Intelligent Systems," \textit{IEEE Commun. Mag.}, vol. 59, no. 6, pp. 96-102, June 2021, doi: 10.1109/MCOM.001.2000604.

\bibitem{ref15}
D. Gündüz et al., ``Beyond Transmitting Bits: Context, Semantics, and Task-Oriented Communications," \textit{IEEE J. Sel. Areas Commun.}, vol. 41, no. 1, pp. 5-41, Jan. 2023, doi: 10.1109/JSAC.2022.3223408.

\bibitem{ref16}
A. Ali, K. Mahmoud and M. Lehtonen, ``Multiobjective Photovoltaic Sizing With Diverse Inverter Control Schemes in Distribution Systems Hosting EVs," \textit{IEEE Trans. Ind. Inform.}, vol. 17, no. 9, pp. 5982-5992, Sept. 2021, doi: 10.1109/TII.2020.3039246.

\bibitem{ref17}
V. Kekatos, G. Wang, A. J. Conejo and G. B. Giannakis, ``Stochastic Reactive Power Management in Microgrids With Renewables," \textit{IEEE Trans. Power Sys.}, vol. 30, no. 6, pp. 3386-3395, Nov. 2015, doi: 10.1109/TPWRS.2014.2369452.

\bibitem{ref18}
The South California Edison's Distributed Energy Resource Interconnection Map (DERiM). [Online]: \url{https://drpep.sce.com/drpep/}.

\bibitem{ref19}
K. Gupta, S. Sahoo, B. K. Panigrahi, F. Blaabjerg and P. Popovski, ``On the assessment of cyber risks and attack surfaces in a real-time co-simulation cybersecurity testbed for inverter-based microgrids", \textit{Energies}, vol. 14, no. 16, pp. 4941, Aug. 2021.

\bibitem{ref20}
M. Elsayed, M. Erol-Kantarci, B. Kantarci, L. Wu and J. Li, ``Low-Latency Communications for Community Resilience Microgrids: A Reinforcement Learning Approach," \textit{IEEE Trans. Smart Grid}, vol. 11, no. 2, pp. 1091-1099, March 2020, doi: 10.1109/TSG.2019.2931753.

\bibitem{ref21}
X. Jiang, J. Zhang, B. J. Harding, J. J. Makela and A. D. Domı´nguez-Garcı´a, ``Spoofing GPS Receiver Clock Offset of Phasor Measurement Units," \textit{IEEE Trans. Power Sys.}, vol. 28, no. 3, pp. 3253-3262, Aug. 2013, doi: 10.1109/TPWRS.2013.2240706.

\bibitem{ref22}
E. Shereen, R. Ramakrishna and G. Dán, ``Detection and Localization of PMU Time Synchronization Attacks via Graph Signal Processing," \textit{IEEE Trans. Smart Grid}, vol. 13, no. 4, pp. 3241-3254, July 2022, doi: 10.1109/TSG.2022.3150954.

\bibitem{ref23}
S. Sahoo, Y. Yang and F. Blaabjerg, ``Resilient Synchronization Strategy for AC Microgrids Under Cyber Attacks," \textit{IEEE Trans. Power Electron.}, vol. 36, no. 1, pp. 73-77, Jan. 2021, doi: 10.1109/TPEL.2020.3005208.

\bibitem{ref24}
F. L. Lewis, H. Zhang, K. Hengster-Movric, and A. Das, \textit{Cooperative Control of Multi-Agent Systems: Optimal and Adaptive Design Approaches}. New York, NY, USA: Springer-Verlag, 2014.

\bibitem{A2}
K. Gupta, S. Sahoo and B. K. Panigrahi, ``Delay-Aware Semantic Sampling in Power Electronic Systems," \textit{IEEE Trans. Smart Grid}, doi: 10.1109/TSG.2023.3339707 \textit{(Early access)}.

\bibitem{RF23}
L. Sheng, G. Lou, W. Gu, S. Lu, S. Ding and Z. Ye, ``Optimal Communication Network Design of Microgrids Considering Cyber-Attacks and Time-Delays," \textit{IEEE Trans. Smart Grid}, vol. 13, no. 5, pp. 3774-3785, Sept. 2022, doi: 10.1109/TSG.2022.3169343.

\bibitem{RF25}
T. Yang, Y. He and G. -P. Liu, ``Distributed Voltage Restoration of AC Microgrids Under Communication Delays: A Predictive Control Perspective," \textit{IEEE Trans. Circuits Syst. I, Reg. Papers}, vol. 69, no. 6, pp. 2614-2624, June 2022, doi: 10.1109/TCSI.2022.3163204.

\bibitem{RF26}
A. Mohammad Saber, A. Youssef, D. Svetinovic, H. H. Zeineldin and E. F. El-Saadany, ``Anomaly-Based Detection of Cyberattacks on Line Current Differential Relays," \textit{IEEE Trans. Smart Grid}, vol. 13, no. 6, pp. 4787-4800, Nov. 2022, doi: 10.1109/TSG.2022.3185764.

\end{thebibliography}
\end{document}